\documentclass[journal]{IEEEtran}

\usepackage{threeparttable}
\usepackage{hyperref}
\usepackage{url}
\usepackage{amsmath,amssymb}
\usepackage{enumerate}

\newcommand{\beq}{\begin{equation}}
\newcommand{\beql}[1]{\begin{equation}\label{#1}}
\newcommand{\eeq}{\end{equation}}

\usepackage{etoolbox}
\IEEEoverridecommandlockouts
\usepackage{cite}
\usepackage{amsmath,amssymb,amsfonts}

\usepackage{tkz-euclide,tikzscale}
\usetikzlibrary{arrows,circuits.ee.IEC}
\tikzset{ac source/.style={
  circuit symbol lines,
  circuit symbol size = width 2 height 2,
  shape = generic circle IEC,
  /pgf/generic circle IEC/before background={
    \pgfpathmoveto{\pgfpoint{-0.8pt}{0pt}}
    \pgfpathsine{\pgfpoint{0.4pt}{0.4pt}}
    \pgfpathcosine{\pgfpoint{0.4pt}{-0.4pt}}
    \pgfpathsine{\pgfpoint{0.4pt}{-0.4pt}}
    \pgfpathcosine{\pgfpoint{0.4pt}{0.4pt}}
    \pgfusepath{stroke}
  },
  transform shape
}}

\usepackage{graphicx}
\usepackage{textcomp}
\usepackage{xcolor}
\usepackage{graphicx}      
\usepackage{bm}
\usepackage{epsfig}
\usepackage{epstopdf}
\usepackage{float}
\usepackage{mathrsfs}
\usepackage{xtab}
\usepackage{color}
\usepackage{ifthen}
\usepackage{setspace}
\usepackage{algorithm}
\usepackage{algorithmic}
\usepackage{dsfont}
\usepackage{booktabs}
\usepackage{verbatimbox}
\usepackage{subfig}
\usepackage{ntheorem}

\newtheorem{theorem}{Theorem}

\newtheorem{corollary}{Corollary}

\usepackage[font=sc,labelsep=period]{caption}
\usepackage{multirow}
\usepackage{color}
\usepackage{mathrsfs}
\usepackage{bm}
\usepackage{url}
\usepackage[spaces,hyphens]{xurl}
\begin{document}

	\title{Multi-Interval Rolling-Window Joint Dispatch and Pricing of Energy and Reserve under Uncertainty}
	
	\author{Jiantao Shi,
		Ye Guo,\IEEEmembership{~Senior~Member,~IEEE,} Wenchuan Wu,\IEEEmembership{~Fellow,~IEEE,}
		and Hongbin Sun,\IEEEmembership{~Fellow,~IEEE}
		\thanks{Jiantao Shi, Ye Guo, Wenchuan Wu and Hongbin Sun are with the Smart Grid and Renewable Energy Laboratory, Tsinghua-Berkeley Shenzhen Institute, Shenzhen, 518071 China. Wenchuan Wu and Hongbin Sun are also affiliated with the State Key Laboratory of Power Systems, Department of Electrical Engineering, Tsinghua University, Beijing, 100084 China.}}

	    \makeatletter
\patchcmd{\@maketitle}
  {\addvspace{0.5\baselineskip}\egroup}
  {\addvspace{-0.9\baselineskip}\egroup}
  {}
  {}
\makeatother
	
	
	\maketitle
	
    \begin{abstract}
    In this paper, the intra-day multi-interval rolling-window joint dispatch and pricing of energy and reserve is studied under increasing volatile and uncertain renewable generations. A look-ahead energy-reserve co-optimization model is proposed for the rolling-window dispatch, where possible contingencies and load/renewable forecast errors over the look-ahead window are modeled as several scenario trajectories, while generation, especially its ramp, is jointly scheduled with reserve to minimize the expected system cost considering these scenarios. Based on the proposed model, marginal prices of energy and reserve are derived, which incorporate shadow prices of generators' individual ramping capability limits to eliminate their possible ramping-induced opportunity costs or arbitrages. We prove that under mild conditions, the proposed market design provides dispatch-following incentives to generators without the need for out-of-the-market uplifts, and truthful-bidding incentives of price-taking generators can be guaranteed as well. Some discussions are also made on how to fit the proposed framework into current market practice. These findings are validated in numerical simulations.
	\end{abstract}
	
	\IEEEpeerreviewmaketitle

\section{Introduction}

\subsection{Backgrounds}

Climate change caused by excessive carbon emissions puts urgent demands on the transformation of the energy structure towards more renewable generations. However, unlike conventional generators with fuel inventories, power outputs of renewable generators may change rapidly with time and cannot be accurately forecasted. These two properties are often described as the \textit{volatility} and \textit{uncertainty} of renewables. Commonly, they can be handled by \textit{inter-temporal} generation ramp scheduled in the energy market and \textit{intra-temporal} operating reserve procured in the ancillary service (AS) market, respectively. Ramp and reserve activation both require instant changes in controllable generators' outputs, therefore they share generators' physical ramping capabilities, i.e., how fast generators' outputs are able to change, which leads to their ramping-induced coupling. Besides, energy and reserve are also coupled in generation capacity and transmission capacity limits. Therefore, with the increasing renewable penetration in modern power systems, it is important to enhance the multi-interval market scheduling of energy and reserve.

\subsection{Related works}

As recently reported in \cite{TLMP1,TLMP2,Cong2022PSCC}, it is beneficial for the independent system operator (ISO) to implement the multi-interval \textit{rolling-window} dispatch when facing the increasing ramp need and forecast uncertainty brought by high-share renewables. In the rolling-window dispatch, to clear the market for an interval that the system is immediately going to enter, the ISO runs a forecast-based optimization for a look-ahead scheduling window, which starts with that immediate interval and extends several followed ones into the future. From such look-ahead scheduling, only the result for the immediate interval is binding, while those for future ones are just advisory. This is different from the \textit{one-shot} dispatch, where the results from one multi-interval optimization are simultaneously binding for all intervals, just like what many ISOs do for their day-ahead scheduling. Moreover, to set aside reliability margins, a fixed reserve requirement, which is typically selected as the capacity of the largest online unit or a percentage of total loads, can also be incorporated to jointly dispatch energy and reserve \cite{ReserveReview2019}. 

Many industrial practices are in line with above methodologies. For example, in NYISO, a look-ahead co-optimization is run every five minutes to financially and physically schedule energy and AS\cite{NYISOmanual}. While in CAISO, the Fifteen-minute market (FMM) is conducted every quarter-hour to determine financially and physically binding AS schedules along with financially binding energy settlements, and then the fixed AS plans are inputted into the five-minute real-time dispatch to arrange physical energy productions at a smaller timescale\cite{CAISOtariff}. At the same time, to enhance the deliverability of reserve, NYISO specifies separate requirements for three different sets of load zones\cite{NYISOZonalReserve}, while CAISO does this for eight AS sub-regions\cite{CAISOtariff}.

Some challenging issues, though, arise from current market practices. One question is how to price the rolling-window dispatch, and a natural intuition is the rolling-window extension of the locational marginal pricing (R-LMP). Unfortunately, as shown in \cite{TLMP1}, ramping-limited generators may suffer opportunity costs at the binding interval under R-LMP and expect to be compensated during the following advisory ones, but such compensations may never be realized as the look-ahead window rolls forward, thus inducing the \textit{missing-money problem} of generators, as well as unavoidable out-of-the-market uplifts to them thereby. To address this problem, some ISOs purchase ramping capabilities from generators as public products and compensate them, including the Ramp Capability Product in MISO\cite{MISOFRP} and the Flexible Ramping Product in CAISO\cite{CAISOFRP}. However, the actual performance of these ramping products is not satisfactory due to their fairly low or even zero prices during most historical time\cite{FRPcons2019}.

Besides the energy pricing, open questions also lie in the reserve market design, like the empirical reserve requirement and the inflexible reserve zone partition, which may not be able to address renewable power uncertainties or guarantee the deliverability of reserve. Moreover, possible generation re-dispatch cost from reserve activation is not considered in the current merit-order-based dispatch that only accounts for the reserve bid-in cost, and methodologies of reserve cost allocation with solid theoretical basis are still lacking. Some of these topics are of high research priority as reported by the National Renewable Energy Laboratory (NREL) \cite{ResearchPrior2021}. 

Many studies aim to handle these open questions. On one hand, to address the incentive distortion in R-LMP and reduce ad hoc ramping-induced uplifts, several novel uniform pricing schemes have been proposed, e.g., the multi-settlement LMP in \cite{zhao2020} which conducts multiple sequential look-ahead schedules and deviation-based settlements for each interval as an extension of current two-settlement system, and the price-preserving multi-interval pricing (PMP) in \cite{PMPHogan2016} which adopts separate models for dispatch and pricing to minimize uplifts under past settled prices. However, \cite{TLMP1} establishes that should generators be incentivized to follow the rolling-window dispatch, uplifts are somehow unavoidable for any uniform pricing, which make the overall settlement discriminatory.

To avoid such out-of-the-market discrimination, the temporal LMP (TLMP) is proposed in a two-part paper \cite{TLMP1,TLMP2} as a non-uniform pricing mechanism, such that the inevitable discriminatory payment becomes in-market. Specifically, TLMP of each generator includes the public LMP and its private ramping shadow price, and the authors prove that TLMP provides dispatch-following and truthful-bidding incentives to generators under the rolling-window dispatch with no need for uplifts. \cite{Cong2022PSCC} further extends TLMP into a stochastic form.

Although these studies provide valuable methods of rolling-window energy pricing, reserve is somehow absent in their contexts. Some complementary efforts, on the other hand, have been made to fix the reserve-related shortcomings of current market practices, e.g., advanced probabilistic methods of reserve requirement selection\cite{nonpara3,nonpara4}, and deliverability-enhancing methodologies including dynamic reserve zone partitions \cite{area3} and post-event reserve response policies \cite{ResponseSetHedman2}. 

In addition to these patches for the current deterministic market design, there have also been some attempts to apply probabilistic optimization techniques on the energy-reserve joint dispatch, e.g., the robust \cite{UMP} or adjustable robust \cite{AdRobust2} optimization, the chance-constrained optimization \cite{Yury2020}, and the scenario-based optimization\cite{readjust1,SHI2021PESGM,reserve_LMP_1,shi2020scenariooriented} that is also adopted in this paper. In \cite{readjust1,reserve_LMP_1,SHI2021PESGM}, the locational marginal pricing of reserve is developed, and such uniform pricing is further extended to the whole of reserve and re-dispatch in \cite{shi2020scenariooriented}. Furthermore, authors of \cite{shi2020scenariooriented} establish individual rationality and cost recovery of generators for each scenario and revenue adequacy of the ISO in expectation. However, existing scenario-based co-optimization efforts adopt either the single-interval dispatch e.g. \cite{readjust1,SHI2021PESGM,shi2020scenariooriented} or the multi-interval one-shot dispatch e.g. \cite{reserve_LMP_1}, while their generalizations to the multi-interval rolling-window setting are not properly addressed yet.

In a word, although many efforts have been made for the market clearing of energy and reserve, no existing work has adequately handled all their couplings and properly scheduled their multi-interval joint dispatch and pricing in a rolling-window manner.

\subsection{Contributions and organizations}

In this paper, we investigate the multi-interval rolling-window co-optimization and pricing of energy and reserve with the scenario-oriented approach, as well as market properties therein. The main contributions of this work are threefold:

(1) A rolling-window co-optimization model is proposed. Possible contingencies and load/renewable forecast errors over the look-ahead window are modeled as several scenario trajectories, and they are properly addressed by the coordinated procurement of energy and reserve from generators, leading to the minimized expected system total cost.

(2) The rolling-window pricing of energy and reserve is developed following the marginal pricing principle, which provides fundamental insights into their \textit{homo- and heterogeneity}. For generators, the Lagrangian multipliers associated with their ramping limits are incorporated into their energy and reserve prices to eliminate possible ramping-induced underpayments or arbitrages via an in-market way.

(3) Dispatch-following incentives of generators are guaranteed without the need for any out-of-the-market uplift, as well as their cost recovery. Truthful-bidding incentives of price-taking generators are also proven.

The rest of this paper is organized as follows. The co-optimization is formulated in Section II. The pricing mechanism is established in Section III. With the proposed dispatch and pricing, dispatch-following and truthful-bidding incentives of generators are studied in Section IV. Some discussions are made in Section V. Case studies are presented in Section VI. Section VII concludes this paper. Designated symbols are listed in the Appendix A. Note that \textbf{all vectors and matrices are in bold}.

\section{Problem Modeling}

\subsection{Time line and probability setting}

\begin{figure}[t]
	\vspace{-0.30cm}
		\centering
	\vspace{-0.7cm}
		\includegraphics[width=3.6in]{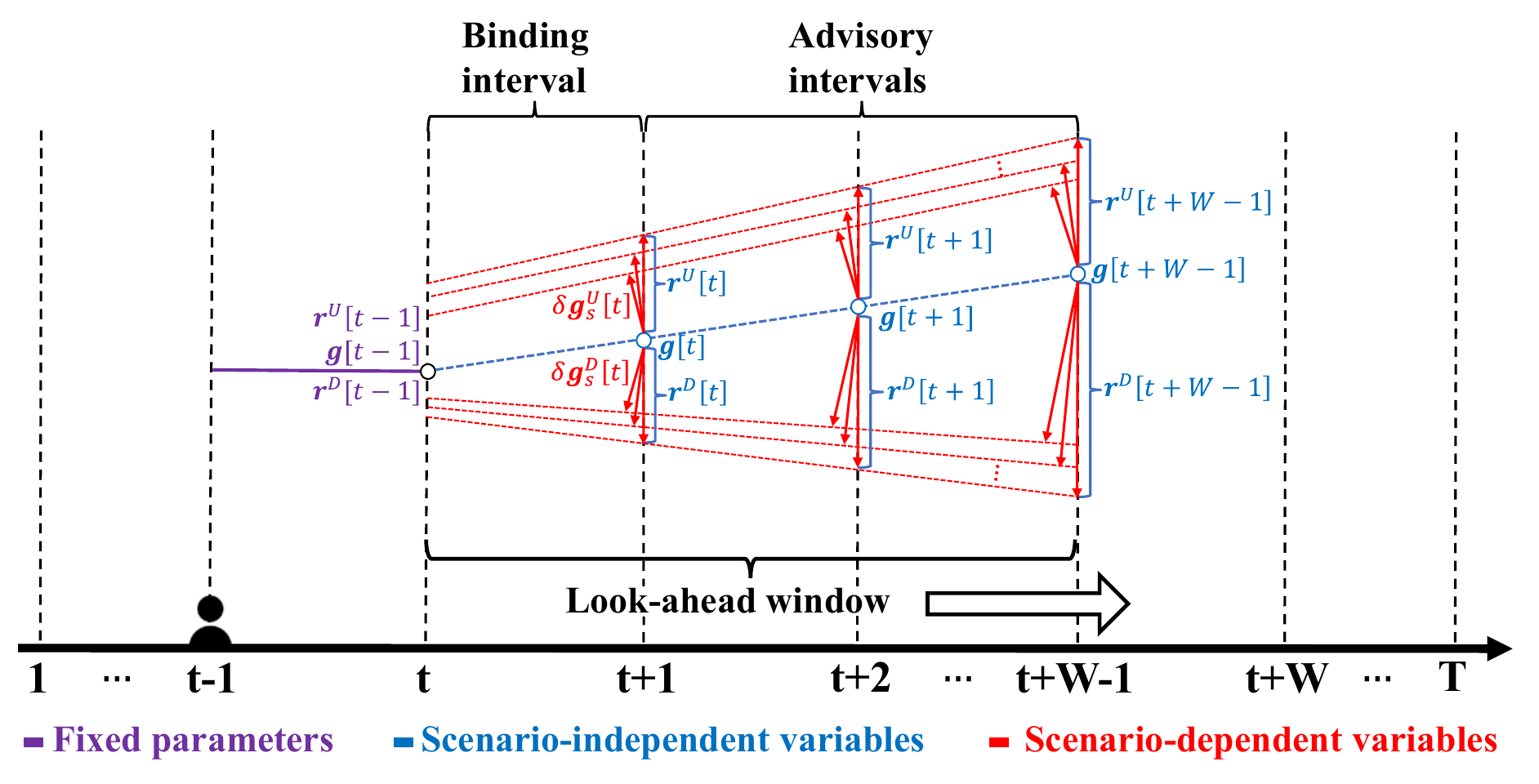}
    	\vspace{-0.55cm}
    \captionsetup{font=footnotesize,singlelinecheck=false}
		\caption{ISO's look-ahead co-optimization over $\mathscr{H}_t$, which produces binding dispatch quantities and prices for the immediate interval $t$ and advisory results for other intervals in $\mathscr{H}_t$ based on the system nominal-operation state (the blue line) and scenario trajectories of possible contingencies and load/renewable forecast errors (red lines)}
		\label{time sequence}
    \vspace{-0.60cm}
	\end{figure}

We consider the intra-day multi-interval joint scheduling of energy and reserve. We assume that $T$ unit-length intervals are involved in the entire scheduling period $\mathscr{H}$ being considered, i.e., $\mathscr{H} = \{1,\cdots,T\}$, and each interval $t$ covers the time duration $[t,t+1)$. To make better use of the continuously updated forecast as time passes, the market over $\mathscr{H}$ is cleared by a sequence of look-ahead co-optimizations, each covers a look-ahead window of $W$ intervals. For example, as shown in Fig. \ref{time sequence}, when the ISO stands at $t-1$, it conducts a co-optimization over the look-ahead window $\mathscr{H}_t=\{t,\cdots, t+W-1\}$ to produce binding dispatch quantities and prices for the immediate interval $t$ and advisory results for other intervals in $\mathscr{H}_t$. For cases where the remaining period is shorter than $W$, the co-optimization is conducted over the \textit{shrinking} look-ahead window $\mathscr{H}_t=\{t,\!\cdots, \!T\}$, from $t$ to the end of $\mathscr{H}$.

Subsequently, when the ISO actually enters interval $t$, on one hand, the ISO implements the binding result for $t$ into real-time system operations, and on the other hand, takes that result as an input for the co-optimization over the next look-ahead window $\mathscr{H}_{t+1}=\{t+1,\cdots, t+W\}$ to clear the market for $t+1$. As time goes by, the look-ahead window rolls through the scheduling period $\mathscr{H}$ along with evolving binding intervals, thus producing energy and reserve schedules over $\mathscr{H}$ in a rolling-window manner. In spirit, such a scheduling process is consistent with the market practice in NYISO \cite{NYISOmanual} and CAISO \cite{CAISOtariff} as described in the Introduction.

Specifically, the co-optimization for each look-ahead window $\mathscr{H}_t$ schedules generation dispatch for system nominal operation, and simultaneously prepares reserve for possible outages and forecast errors. As the benchmark for the nominal dispatch, the forecast power path of net load (including loads and renewables) over $\mathscr{H}_t$ is represented by a blue line in Fig. \ref{time sequence}. And for any interval $\tau$ in $\mathscr{H}_t$, the ISO optimizes energy $\bm{g}[\tau]$ from generators to supply that forecast net load under the contingency-absent state. 

Meanwhile, the ISO's look-ahead forecast for $\mathscr{H}_t$ is not perfect when standing ahead at $t-1$, and we assume that possible contingencies and net load forecast errors over $\mathscr{H}_t$ are modeled as a set of scenario trajectories $\mathcal{S}_t$ and represented by those red lines in Fig. \ref{time sequence}, each of which considers a path of net load forecast error and a sequence of contingency realization over $\mathscr{H}_t$. To cope with possible realizations of any scenario $s$ in system real-time operations, remedial upward and downward re-dispatch $\bm{\delta g}^U_s[\tau], \bm{\delta g}^D_s[\tau]$ of generators are expected, and we endogenously set their reserve provision $\bm{r}^U[\tau],\bm{r}^D[\tau]$ as the biggest range of their generation re-dispatch among all possible scenarios, such that the exogenous reserve requirement is eliminated. In the next subsection, the corresponding co-optimization is formulated in detail.

\subsection{Model formulation}

In this paper, the following assumptions are adopted to formulate our energy-reserve co-optimization model:

i) The cost functions of each generator for providing energy, reserve, and re-dispatch are assumed to be linear.

ii) Renewable generations are treated as negative loads.

iii) A standard DC power flow model with line-flow constraints is adopted. Power losses on transmission lines are ignored.
	
iv) Possible line contingencies and load forecast errors over each look-ahead window $\mathscr{H}_t$ are modeled as a set of scenario trajectories $\mathcal{S}_t$ and handled by reserve, while generator outages are ignored to avoid generation heterogeneity from different outage probabilities.
		
v) The probability of each scenario is assumed to be given. 

Validations and implications of above assumptions will be revisited in Section \ref{Sec:Discussion}. Accordingly, the co-optimization model for $\mathscr{H}_t$ can be formulated as:
	\vspace{-4pt}
	\begin{flalign}
	\label{obj}
    &{\rm minimize} \ F^t(g_{i\tau},r^U_{i\tau},r^D_{i\tau}, \delta g^U_{i\tau,s}, \delta g^D_{i\tau,s}, \delta d_{l\tau,s})=\notag&\\
    &\sum\limits_{\tau \in \mathscr{H}_t}\sum\limits_{i \in N_G} \Big( C^g_{i\tau}g_{i\tau}+C^U_{i\tau}r^U_{i\tau}+C^D_{i\tau}r^D_{i\tau}\Big)\notag&\\
    &+ \!\sum\limits_{s \in \mathcal{S}_t}\epsilon_s\sum\limits_{\tau \in \mathscr{H}_t}\sum\limits_{i \in N_G}\Big(C^g_{i\tau}\delta g^U_{i\tau,s}\!-\!C^g_{i\tau}\delta g^D_{i\tau,s}\Big)\notag&\\
    &+ \!\sum\limits_{s \in \mathcal{S}_t}\epsilon_s\sum\limits_{\tau \in \mathscr{H}_t}\sum\limits_{l \in N_D}C^L_{l\tau}\delta d_{l\tau,s},&\\
    &\mbox{subject to $\forall \tau \in \mathscr{H}_t$:}\notag&\\
	&\mbox{Ex-ante network constraints for the nominal dispatch:}\notag&\\
	\label{base balance}
	&\lambda_{\tau,0}:\sum\limits_{i \in N_G}\! g_{i\tau}\!=\!\sum\limits_{l \in N_D}\! \hat{d}_{l\tau},&\\
	\label{base pf}
	&\bm{\phi}_{\tau,0}:\bm{S}_{G}\cdot\bm{g}[\tau]\!-\bm{S}_{D}\cdot\hat{\bm{d}}[\tau]\!\leq\! \bm{f},&\\
	&\mbox{Ex-post network constraints after re-dispatch $\forall s \in \mathcal{S}_{t}$:}\notag&\\
	\label{non-base scenario balance}
    &\lambda_{\tau,s}:\sum\limits_{i \in N_G} \!(g_{i\tau}+\!\delta g^U_{i\tau,s}\!-\!\delta g^D_{i\tau,s})\!=\!\sum\limits_{l \in N_D}\!(\hat{d}_{l\tau}\!+\!\hat{\xi}_{l\tau,s}\!-\!\delta d_{l\tau,s}),&\\
	&\bm{\phi}_{\tau,s}:\bm{S}_{G,\tau,s}\cdot(\bm{g}[\tau]+\bm{\delta g}^U_{s}[\tau]-\bm{\delta g}^D_{s}[\tau])&\notag\\
	\label{non-base scenario pf}
	&\ \ \ \ \ \ \ -\bm{S}_{D,\tau,s}\!\cdot\!\big(\hat{\bm{d}}[\tau]\!+\!\hat{\bm{\xi}}_s[\tau]\!-\!\bm{\delta d}_{s}[\tau]\big)\!\leq\! \bm{f}_{\tau,s},&\\
	&\mbox{Re-dispatch constraints $\forall s \in \mathcal{S}_{t}, \ \forall i \in N_G$:}&\notag\\
	\label{dr1 range}
    &(\underline{\alpha}_{i\tau,s},\overline{\alpha}_{i\tau,s}): 0 \leq \delta g^U_{i\tau,s} \leq r^U_{i\tau},&\\
    \label{dr2 range}
    &(\underline{\beta}_{i\tau,s},\overline{\beta}_{i\tau,s}): 0 \leq \delta g^D_{i\tau,s} \leq r^D_{i\tau},&\\
    &\mbox{Load shedding constraints $\forall s \in \mathcal{S}_{t},\ \forall l \in N_D$:}&\notag\\
	\label{shedding non-base scenario}
	&(\underline{\gamma}_{l\tau,s},\overline{\gamma}_{l\tau,s}): 0 \leq \delta d_{l\tau,s} \leq \hat{d}_{l\tau}+\hat{\xi}_{l\tau,s},&\\
    &\mbox{Generation, reserve, and ramping capability limits $\forall i \in N_G$:}&\notag\\
    \label{binding capacity}	
	&(\underline{\upsilon}_{i\tau},\overline{\upsilon}_{i\tau}):\underline{G}_{i\tau} \leq g_{i\tau}-r^D_{i\tau},g_{i\tau}+r^U_{i\tau} \leq \overline{G}_{i\tau},&\\
	\label{reserve capability}
    &(\underline{\rho}^D_{i\tau},\overline{\rho}^D_{i\tau},\underline{\rho}^U_{i\tau},\overline{\rho}^U_{i\tau}):0 \leq r^D_{i\tau} \leq R^D_{i\tau},0 \leq r^U_{i\tau} \leq R^U_{i\tau},&\\
	\label{binding inter-temporal down}
	&\underline{\mu}_{i(\tau-1)}:-\underline{r}_{i\tau} \leq g_{i\tau}-g_{i(\tau-1)}-r^D_{i\tau}-r^U_{i(\tau-1)},&\\
	\label{binding inter-temporal up}
	&\overline{\mu}_{i(\tau-1)}: g_{i\tau}-g_{i(\tau-1)}+r^U_{i\tau}+r^D_{i(\tau-1)} \leq \overline{r}_{i\tau}.&
	\end{flalign}
	The objective function (\ref{obj}) consists of three parts over $\mathscr{H}_t$: i) the bid-in cost of energy $g_{i\tau}$, upward $r^U_{i\tau}$ and downward $r^D_{i\tau}$ reserve in the first row where $i$ is the index of generators, ii) the expected cost for ex-post upward and downward re-dispatch $\delta g^U_{i\tau,s},\delta g^D_{i\tau,s}$ among all possible scenarios $\mathcal{S}_t$ in the second row where $\epsilon_s$ is the probability of any scenario $s$,\footnote{In this paper, we do not distinguish the cost coefficients of generation and its re-dispatch. Should some generators suffer additional costs for fast ramp, their cost coefficients for re-dispatch can be differentiated from $C^g_{i\tau}$ as in \cite{shi2020scenariooriented}. Nevertheless, such settings do not influence the qualitative analysis in this paper.} and iii) the expected cost for ex-post load shedding $\delta d_{l\tau,s}$ among $\mathcal{S}_t$ in the third row where $l$ is the index of loads.
	
    For the constraint part, (\ref{base balance})-(\ref{base pf}) are the ex-ante energy balancing and transmission capacity constraints for the nominal dispatch at each interval $\tau$ in $\mathscr{H}_t$, respectively, where $\hat{d}_{l\tau}$ is the forecast power of load $l$ at $\tau$, and $\bm{g}[\tau],\hat{\bm{d}}[\tau]$ are vector forms of nominal generation and forecast consumption from all resources, respectively. $\bm{S}_G$ and $\bm{S}_D$ are the shift-factor matrices correspond to generators and loads, respectively, and $\bm{f}$ is the transmission capacity vector. (\ref{non-base scenario balance})-(\ref{non-base scenario pf}) are the ex-post network constraints after the realization of any scenario $s$ at $\tau$ with i) line outages reflected in $\bm{S}_{G,\tau,s},\bm{S}_{D,\tau,s},\bm{f}_{\tau,s}$ different from their nominal values $\bm{S}_G,\bm{S}_D,\bm{f}$, ii) load forecast errors $\hat{\bm{\xi}}_s[\tau]$ from their forecast values, and iii) remedial generation re-dispatch and load shedding. These constraints capture the influence of possible re-adjustments on network power flow, thus guaranteeing the scenario-wise deliverability of reserve without the need for empirical reserve zone partition. (\ref{dr1 range})-(\ref{dr2 range}) link ex-ante reserve procurement with possible ex-post re-dispatch, and (\ref{shedding non-base scenario}) limits load shedding in all scenarios.
	
\begin{figure}[t]
		\centering
	\vspace{-0.70cm}
		\includegraphics[width=2.9in]{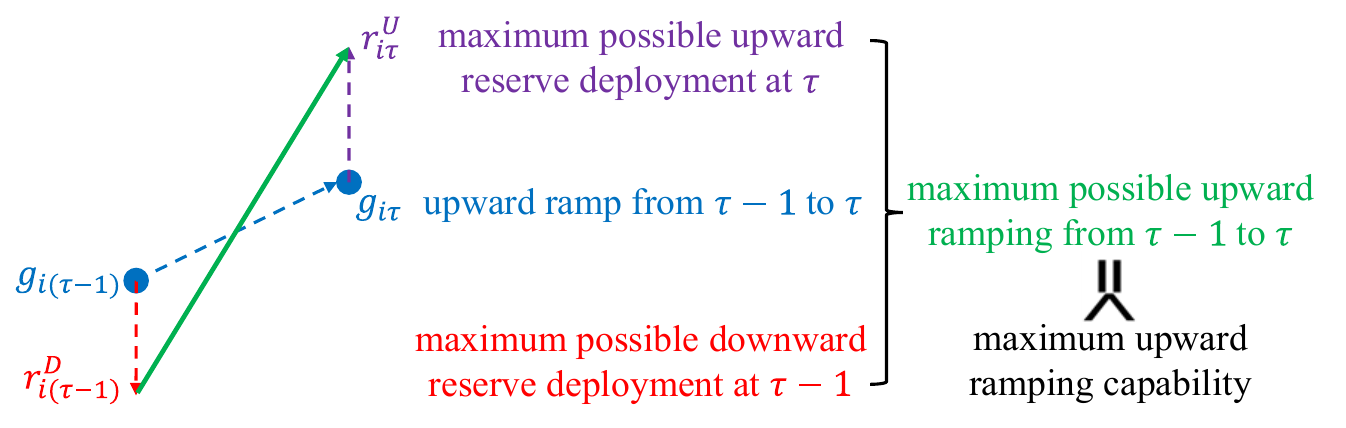}
    \captionsetup{font=footnotesize,singlelinecheck=false}
		\caption{Description of the upward ramping limit (\ref{binding inter-temporal up})}
		\label{fig:ramping depict}
    \vspace{-0.60cm}
	\end{figure}

	Moreover, (\ref{binding capacity})-(\ref{reserve capability}) enforce generators' capacity and reserve capability limits while (\ref{binding inter-temporal down})-(\ref{binding inter-temporal up}) represent ramping ones. Note that in the extreme case, a generator may have opposite re-dispatch directions for adjacent time intervals as illustrated in Fig. \ref{fig:ramping depict}, thus its ramp and reserve should satisfy (\ref{binding inter-temporal down})-(\ref{binding inter-temporal up}).

    Note that the solution of the proposed model is much more time-consuming than traditional ones, especially when a large amount of scenarios are considered. Many decomposition methods are applicable to accelerate the solution process e.g. \cite{Nasri2016onBD,CRE}. In this paper, we do not focus on the detailed solution algorithm, but properties of such solution results.
    
    By sequentially solving the proposed model $F^t(\cdot)$ over $\mathscr{H}$ in a rolling-window manner, the energy and reserve dispatch for each generator $i$ can be defined as
    \begin{equation}
    \label{rolling window dispatch}
    \setlength\abovedisplayskip{0.5pt}
	\setlength\belowdisplayskip{0.5pt}
    \widetilde{\bm{P}}_i=\{g^*_{it},r^{U*}_{it},r^{D*}_{it}, \forall t \in \mathscr{H}\}, 
    \end{equation}
    and the associated pricing mechanism is developed in the next section.

\section{Pricing Mechanism}
The proposed pricing mechanism is developed based on the marginal pricing principle. For any interval $t$, binding settlement prices of energy and reserve are derived from the co-optimization $F^t(\cdot)$ over the corresponding look-ahead window $\mathscr{H}_t$. Note that energy and reserve from generators are decision variables in $F^t(\cdot)$, whose marginal costs cannot be directly derived. To solve this problem, for each generator $i$, we fix its energy and reserve $g_{it},r^U_{it},r^D_{it}$ for $t$ at their optimums $g^*_{it},r^{U*}_{it},r^{D*}_{it}$, convert these variables to parameters in $F^t(\cdot)$, and consider the rest of the market. Thus we can get a \textit{parameterized} model
	\begin{equation}
	\setlength\abovedisplayskip{0pt}
	\setlength\belowdisplayskip{0pt}
	\underset{x_{-it} \in \mathcal{X}_{-it}}{\rm minimize} \ F^t_{-it}(x_{-it}),\notag
	\end{equation}
where $x_{-it}$ represents all decision variables of $F^t(\cdot)$ except for $g_{it},r^U_{it},r^D_{it}$ of generator $i$ at $t$, $F^t_{-it}(\cdot)$ represents the system expected cost (\ref{obj}) over $\mathscr{H}_t$ excluding the energy and reserve bid-in cost of generator $i$ at $t$, and $\mathcal{X}_{-it}$ includes all the constraints in (\ref{base balance})-(\ref{binding inter-temporal up}) except for the intra-temporal internal generation capacity and reserve capability limits (\ref{binding capacity})-(\ref{reserve capability}) of generator $i$ at $t$. Then we can derive prices for generator $i$ at $t$ by evaluating the sensitivity of the optimal objective function $F^t_{-it}(x^*_{-it})$. Namely, its energy marginal price at $t$ is
\vspace{-0.15cm}
	\begin{align}
	\label{Gen energy price}
	\pi^g_{it}=&-\frac{\partial}{\partial g_{it}} F^t_{-it}(x_{-it}^*)\notag\\
	=&\lambda^*_{t,0}-\bm{S}^T_G(\cdot,m_i)\bm{\phi}^*_{t,0}+\sum\limits_{s \in \mathcal{S}_{t}}\Big(\lambda^*_{t,s}-\bm{S}^T_{G,t,s}(\cdot,m_i)\bm{\phi}^*_{t,s}\Big)\notag\\
	&+\Big(\overline{\mu}^*_{it}-\underline{\mu}^*_{it}\Big)-\Big(\overline{\mu}^*_{i(t-1)}-\underline{\mu}^*_{i(t-1)}\Big)\notag\\
	=&\omega^*_{m_it,0}+\sum\limits_{s \in \mathcal{S}_{t}}\omega^*_{m_it,s}+\Delta \mu^*_{it}- \Delta \mu^*_{i(t-1)},
	\end{align}
	where $m_i$ is the index of the bus connecting with generator $i$. This energy price consists of two components: 
	\\(i) The second row of (\ref{Gen energy price}), denoted by $\omega^*_{m_it,0}+\sum \omega^*_{m_it,s}$, represents components of energy balancing and congestion for both the nominal operation $\omega^*_{m_it,0}$ and possible non-base scenarios $\sum \omega^*_{m_it,s}$, which follows the standard form of lossless LMP;
    \\(ii) The third row represents ramping components. Specifically, $\overline{\mu}^*_{it}-\underline{\mu}^*_{it}$, denoted by $\Delta \mu^*_{it}$, represents the ramping component from time $t$ to $t+1$, which is non-zero if the generator's ramping constraints (\ref{binding inter-temporal down})-(\ref{binding inter-temporal up}) is binding from $t$ to $t+1$. Similarly, $-(\overline{\mu}^*_{i(t-1)}-\underline{\mu}^*_{i(t-1)})$, denoted by $- \Delta \mu^*_{i(t-1)}$, represents the ramping component from $t-1$ to $t$.
	
    We can observe that the energy price $\pi^g_{it}$ has a locational uniform part (i) and a discriminatory part (ii), which provides fundamental insights into the homo- and heterogeneity of energy from generators: generation resources at the same bus are \textit{homogeneous} for their \textit{non-differentiated} contributions to the intra-temporal energy balancing and network congestions, and they are \textit{heterogeneous} for their \textit{differentiated} contributions to the inter-temporal ramp flexibility.
	
    Similarly, its upward and downward reserve marginal prices at $t$ can be derived as
    \begin{equation}
	\label{Gen up res price}
	\setlength\abovedisplayskip{4pt}
	\setlength\belowdisplayskip{0.5pt}
	\pi^U_{it}=-\frac{\partial}{\partial r^U_{it}} F^t_{-it}(x_{-it}^*)=\sum\limits_{s \in \mathcal{S}_{t}} \overline{\alpha}^*_{it,s}-\overline{\mu}^*_{i(t-1)}-\underline{\mu}^*_{it},
	\end{equation}
	    \begin{equation}
	\label{Gen down res price}
	\setlength\abovedisplayskip{0.5pt}
	\setlength\belowdisplayskip{2.5pt}
	\pi^D_{it}=-\frac{\partial}{\partial r^D_{it}} F^t_{-it}(x_{-it}^*)=\sum\limits_{s \in \mathcal{S}_{t}} \overline{\beta}^*_{it,s}-\underline{\mu}^*_{i(t-1)}-\overline{\mu}^*_{it},
	\end{equation}
	For the first component $\sum \overline{\alpha}^*_{it,s}$ in the upward reserve price (\ref{Gen up res price}), if generator $i$'s procured upward reserve \textit{capacity} is fully squeezed into upward re-dispatch in scenario $s$, i.e., $\delta g^U_{it,s}=r^U_{it}$, then the binding upward re-dispatch constraint (\ref{dr1 range}) leads to a non-zero multiplier $\overline{\alpha}^*_{it,s}$ and contributes to this \textit{capacity} component. Note that generators at the same bus may receive different capacity components in (\ref{Gen up res price}) for their different re-dispatch cost coefficients, which make their upward reserve capacities \textit{heterogeneous}. Other terms in (\ref{Gen up res price}) are ramping components just like components (ii) in the generation price (\ref{Gen energy price}) since reserve and ramp share the ramping capabilities of generators. Similar analysis also applies to the downward reserve price (\ref{Gen down res price}). Moreover, according to (\ref{Clearing KKT up reserve})-(\ref{Clearing KKT down reserve}) in the Appendix B, we can interpret that any generator with non-zero upward/downward reserve dispatch quantities at $t$ would definitely receive non-negative upward/downward reserve marginal prices for $t$.
	
	Besides prices for generators, we also derive the energy marginal price of load $l$ at $t$:
    \vspace{-0.15cm}
	\begin{align}
	\setlength{\abovedisplayskip}{0pt}
	\label{Load energy price}
	\pi^d_{lt}=&\frac{\partial}{\partial \hat{d}_{lt}} F^t_{-it}(x_{-it}^*)\notag\\
	=&\lambda^*_{t,0}-\bm{S}^T_D(\cdot,m_l)\bm{\phi}^*_{t,0}+\sum\limits_{s \in \mathcal{S}_{t}}\Big(\lambda^*_{t,s}-\bm{S}^T_{D,t,s}(\cdot,m_l)\bm{\phi}^*_{t,s}\Big)\notag\\
	&+\sum\limits_{s \in \mathcal{S}_{t}} \overline{\gamma}^*_{lt,s}=\omega^*_{m_lt,0}+\sum\limits_{s \in \mathcal{S}_{t}}\omega^*_{m_lt,s}+\sum\limits_{s \in \mathcal{S}_{t}}\overline{\gamma}^*_{lt,s},
	\end{align}
    which includes similar network components while excluding ramping components compared with the generation price in (\ref{Gen energy price}), as well as an additional component $\sum\overline{\gamma}^*_{lt,s}$ brought by the load shedding limit (\ref{shedding non-base scenario}). Typically, a non-zero $\overline{\gamma}^*_{lt,s}$ appears when load $l$ is totally shed in any scenario $s$, which reveals its \textit{differentiated} reliability requirement from other loads and brings in the \textit{heterogeneity} of electrical demands. If ignoring such cases, load energy prices then become locational uniform. 
	
    For the settlement of each interval $t$, each generator is credited for its procured energy, upward reserve, and downward reserve at $t$, respectively settled with (\ref{Gen energy price})-(\ref{Gen down res price}), and each load is charged for its forecast power, settled with (\ref{Load energy price}). Besides, each load also needs to pay for its possible deviations from its forecast power in all scenarios as (assuming $\overline{\gamma}^*_{lt,s}$ to be zero):
	\begin{equation}
	\label{Load fluctuation payment}
	\setlength\abovedisplayskip{0pt}
    \setlength\belowdisplayskip{0pt}
	\Gamma^F_{lt}=\sum\limits_{s \in \mathcal{S}_{t}} \frac{\partial F^t_{-it}(x_{-it}^*)}{\partial \hat{\xi}_{lt,s}} \hat{\xi}_{lt,s}=\sum\limits_{s \in \mathcal{S}_{t}}\omega^*_{m_lt,s}\hat{\xi}_{lt,s}.
	\end{equation}
    As a result, reserve costs can be endogenously distributed to uncertainty sources in a scenario-wise way. 
    
    With the look-ahead window sliding across the scheduling period $\mathscr{H}$, the pricing of energy and reserve for each generator $i$ over $\mathscr{H}$ can be developed in a rolling-window manner as
    \begin{equation}
    \label{rolling window pricing}
    \setlength\abovedisplayskip{1pt}
    \setlength\belowdisplayskip{1pt}
    \widetilde{\bm{\pi}}_i\!=\!\{\pi^g_{it},\pi^U_{it},\pi^D_{it}, \forall t \in \mathscr{H}\}.
    \end{equation}
    Based on the rolling-window dispatch $\widetilde{\bm{P}}_i$ in (\ref{rolling window dispatch}) and pricing $\widetilde{\bm{\pi}}_i$, we study some market properties in the next section. 

\section{Properties}
\label{sec:properties}
In this section, we mainly focus on dispatch-following and truthful-bidding incentives of generators under the proposed rolling-window market-clearing framework.

\subsection{Dispatch-following incentives}
Dispatch-following incentives connote that with given prices from the ISO, any generator can maximize its profit by just following the ISO's dispatch quantities. Specifically, under the proposed market design, after the system has operated throughout the scheduling period $\mathscr{H}$, the energy and reserve clearing prices of any generator $i$ over $\mathscr{H}$ are given by $\widetilde{\bm{\pi}}_i$ in (\ref{rolling window pricing}), then the maximum overall profit that generator $i$ would have received from self-scheduling energy and reserve over $\mathscr{H}$ can be reviewed with its multi-interval profit-maximization:
	\begin{equation}
	\label{obj_Expost_IR_multi}
	\setlength\abovedisplayskip{3pt}
    \setlength\belowdisplayskip{-6pt}
	Q_i(\widetilde{\bm{\pi}}_i)\!=\!\!\!\sum_{t \in \mathscr{H}}(\pi_{it}g_{it}\!+\!\pi^U_{it}\!r^U_{it}\!+\!\pi^D_{it}\!r^D_{it}\!-\!C^g_{it}g_{it}\!-\!C^U_{it}\!r^U_{it}\!-\!C^D_{it}\!r^D_{it}),
	\end{equation}	
	\vspace{-1.2cm}
	\begin{flalign}
    &\underset{\{g_{it},r^U_{it},r^D_{it},t\in\mathscr{H}\}}{\rm maximize}\! Q_i(\widetilde{\bm{\pi}}_i), \mbox{subject to  $\forall t \in \mathscr{H}$:}\notag&\\
	\label{binding capacity 3}	
	&(\underline{\upsilon}'_{it},\overline{\upsilon}'_{it}):\underline{G}_{it} \leq  g_{it} - r^D_{it},g_{it} + r^U_{it}  \leq \overline{G}_{it},&\\
	\label{reserve capability 3}
    &(\underline{\rho}^{D'}_{it},\overline{\rho}^{D'}_{it},\underline{\rho}^{U'}_{it},\overline{\rho}^{U'}_{it}):0 \leq r^D_{it} \leq R^D_{it},0 \leq r^U_{it} \leq R^U_{it},&\\
	 \label{binding inter-temporal up 3}
	&\underline{\mu}'_{i(t-1)}: -\underline{r}_{it} \leq g_{it}-g_{i(t-1)}-r^D_{it}-r^U_{i(t-1)},&\\
	\label{binding inter-temporal down 3}
	&\overline{\mu}'_{i(t-1)}:g_{it}-g_{i(t-1)}+r^U_{it}+r^D_{i(t-1)} \leq \overline{r}_{it}.&
	\end{flalign}    
    If the energy and reserve dispatch $\widetilde{\bm{P}}_i$ from the ISO in (\ref{rolling window dispatch}) optimizes the above profit-maximization model, then the pricing $\widetilde{\bm{\pi}}_i$ is said to provide dispatch-following incentives to generators under $\widetilde{\bm{P}}_i$. On the contrary, if any generator $i$ can be better-off by deviating from $\widetilde{\bm{P}}_i$ and self-scheduling itself, some \textit{lost-of-opportunity-cost} (LOC) uplifts shall be needed to compensate its inadequate revenue:
     
    \begin{equation}
    \label{LOC calculation}
    \setlength\abovedisplayskip{-1pt}
	\setlength\belowdisplayskip{0pt}
LOC(\widetilde{\bm{P}}_i,\widetilde{\bm{\pi}}_i)=Q^*_i(\widetilde{\bm{\pi}}_i)-Q_i(\widetilde{\bm{P}}_i|\widetilde{\bm{\pi}}_i),
    \end{equation}
    where $Q^*_i(\widetilde{\bm{\pi}}_i)$ is the optimal self-scheduling profit solved from (\ref{obj_Expost_IR_multi})-(\ref{binding inter-temporal down 3}), and $Q_i(\widetilde{\bm{P}}_i|\widetilde{\bm{\pi}}_i)$ is the actual profit by following $\widetilde{\bm{P}}_i$ from the ISO. We establish that the proposed pricing mechanism provides dispatch-following incentives to generators without paying them any extra LOC uplift:
    \vspace{-0.15cm}
	\begin{theorem}[Dispatch-following Incentives]
	\ \\ Under assumptions (i)-(v), consider any generator $i$. Over the entire scheduling period $\mathscr{H}$, its energy and reserve dispatch quantities $\widetilde{\bm{P}}_i=\{g^*_{it},r^{U*}_{it},r^{D*}_{it}, \forall t \in \mathscr{H}\}$ and prices $ \widetilde{\bm{\pi}}_i=\{\pi^g_{it},\pi^U_{it},\pi^D_{it}, \forall t \in \mathscr{H}\}$ are solved from the proposed model $F^t(\cdot)$ in a rolling-window manner. Then
	\begin{equation}
	\setlength\abovedisplayskip{1.5pt}
	\setlength\belowdisplayskip{1.5pt}
    \{g^*_{it},r^{U*}_{it},r^{D*}_{it}, \forall t\! \in \! \mathscr{H}\} \!=\!\!\!\underset{\{g_{it},r^{U}_{it},r^{D}_{it}, \forall t  \in \mathscr{H}\}}{\operatorname{argmax}}\!\{Q_i(\widetilde{\bm{\pi}}_i)|(\ref{binding capacity 3})\!-\!(\ref{binding inter-temporal down 3})\!\}.\notag
    \end{equation}
    In another word, $LOC(\widetilde{\bm{P}}_i,\widetilde{\bm{\pi}}_i) = 0$.
	\end{theorem}
	\vspace{-0.15cm}
	This Theorem ensures dispatch-following incentives of generators. Please check the Appendix B for the proof. A consequent corollary is cost recovery of generators with one additional condition:
	\vspace{-0.15cm}
	\begin{corollary}[Cost Recovery]
    \ \\ Under assumptions (i)-(v), for any generator $i$ with zero generation lower bound i.e. $\underline{G}_{it}=0$, by following the ISO's dispatch $\widetilde{\bm{P}}_i$ over $\mathscr{H}$, its total bid-in profit for providing energy and reserve $Q_i(\widetilde{\bm{P}}_i|\widetilde{\bm{\pi}}_i)$ is always non-negative.
	\end{corollary}
	See the Appendix C for the proof. Note that the zero-lower-bound condition is common in the electricity market when considering generators' cost recovery.

\subsection{Truthful-bidding incentives}

Besides the willingness to follow the ISO's dispatch, we also want to see generators' voluntarily truthful bidding. This relies on the additional price-taking assumption, i.e., any generator cannot influence the assigned energy and reserve prices from the ISO, which is common in competitive electricity markets.

With this additional assumption, we further analyze possible bidding behaviors of any price-taking generator $i$ for the entire scheduling period $\mathscr{H}$. We denote its true marginal costs of energy and reserve and true operational parameters of generation, reserve, and ramping capabilities over $\mathscr{H}$ with superscript of $*$ as 
\begin{equation}
    \setlength\abovedisplayskip{0pt}
	\setlength\belowdisplayskip{0pt}
    \bm{\theta}^*_{i}\!=\!\{C^{g*}_{it},C^{U*}_{it},C^{D*}_{it},\overline{G}^*_{it},\underline{G}^*_{it},R^{U*}_{it},R^{D*}_{it},\overline{r}^*_{it},\underline{r}^*_{it},\forall t\! \in\! \mathscr{H}\}.\notag
\end{equation}
However, the generator may not submit above true information to the ISO. Instead, it considers the following strategic bidding problem to maximize its profit over $\mathscr{H}$:
\vspace{-0.2cm}
\begin{align}
	\label{obj_Exante_bidding}
	&\Pi_{i}(\bm{\theta}_{i})=\sum_{t\in \mathscr{H}}\big(\pi^g_{it}g^*_{it}(\bm{\theta}_{i})+\pi^U_{it}r^{U*}_{it}(\bm{\theta}_{i})+\pi^D_{it}r^{D*}_{it}(\bm{\theta}_{i})\big)\notag\\
	&-\!\sum_{t\in \mathscr{H}}\big(C^{g*}_{it}g^*_{it}(\bm{\theta}_{i})+C^{U*}_{it}r^{U*}_{it}(\bm{\theta}_{i})+C^{D*}_{it}r^{D*}_{it}(\bm{\theta}_{i})\big),\\
    &\underset{\{\bm{\theta}_{i}\}}{\rm maximize}\:\: \Pi_{i}(\bm{\theta}_{i}),\: \mbox{subject to }\forall t \in \mathscr{H}:\notag\\
	\label{binding capacity bidding}	
	& \underline{G}^*_{it} \leq g^*_{it}(\bm{\theta}_{i}) - r^{D*}_{it}(\bm{\theta}_{i}),g^*_{it}(\bm{\theta}_{i}) + r^{U*}_{it}(\bm{\theta}_{i})  \leq \overline{G}^*_{it},\\
	\label{binding reserve capability bidding}
    &0 \leq r^{D*}_{it}(\bm{\theta}_{i}) \leq R^{D*}_{it},0 \leq r^{U*}_{it}(\bm{\theta}_{i}) \leq R^{U*}_{it},&\\
    \label{binding inter-temporal up bidding}
	&-\underline{r}^*_{it} \leq g^*_{it}(\bm{\theta}_{i})-g^*_{i(t-1)}(\bm{\theta}_{i})-r^{D*}_{it}(\bm{\theta}_{i})-r^{U*}_{i(t-1)}(\bm{\theta}_{i}),\\
	\label{binding inter-temporal down bidding}
	&g^*_{it}(\bm{\theta}_{i})-g^*_{i(t-1)}(\bm{\theta}_{i})+r^{U*}_{it}(\bm{\theta}_{i})+r^{D*}_{i(t-1)}(\bm{\theta}_{i}) \leq \overline{r}^*_{it},
    \end{align}
    where $\bm{\theta}_{i}$ represents any feasible bid of generator $i$ over $\mathscr{H}$, and $\{g^*_{it}(\bm{\theta}_{i}),r^{U*}_{it}(\bm{\theta}_{i}),r^{D*}_{it}(\bm{\theta}_{i}),\forall t \in \mathscr{H}\}$ with superscript of $*$ are the procured energy and reserve quantities of generator $i$ over $\mathscr{H}$ from the ISO corresponding to any $\bm{\theta}_{i}$. By submitting different $\bm{\theta}_{i}$, generator $i$'s procured quantities would also be different, but the clearing prices $\widetilde{\bm{\pi}}_i=\{\pi^g_{it},\pi^U_{it},\pi^D_{it},\forall t \in \mathscr{H}\}$ from the ISO would never be influenced by the changing bid. Specifically, with $\bm{\theta}_{i}=\bm{\theta}^*_{i}$, generator $i$ is said to bid truthfully for $\mathscr{H}$, and the following Theorem establishes that it is optimal for any price-taking generator to do so:
    \vspace{-0.2cm}
\begin{theorem}[Truthful-bidding Incentives]
	\ \\Under assumptions (i)-(v) and that any generator cannot influence the assigned energy and reserve prices from the ISO, consider any generator $i$. To maximize its profit over $\mathscr{H}$ under the dispatch $\widetilde{\bm{P}}_i$ and pricing $\widetilde{\bm{\pi}}_i$ from the ISO, it is optimal for the generator to reveal its true marginal costs and operational parameters with its bid for $\mathscr{H}$, i.e.,
	\begin{equation}
	\setlength\abovedisplayskip{0pt}
	\setlength\belowdisplayskip{-0.5pt}
    \bm{\theta}^*_{i}=\underset{\{\bm{\theta}_{i}\}}{\operatorname{argmax}}\{\Pi_{i}(\bm{\theta}_{i})|(\ref{binding capacity bidding})-(\ref{binding inter-temporal down bidding})\}.\notag
    \end{equation}
\end{theorem}
\vspace{-0.10cm}
Check the Appendix D for the proof. These properties validate the effectiveness of the proposed market design.

\section{Discussions}
\label{Sec:Discussion}
In this section, we review the assumptions raised in Section II, and make some extended discussions.

\subsection{Revisit of assumptions}

In assumption (i), we adopt linear bid-in cost functions of energy and reserve for generators. This setting is common in the energy market, but necessities of the reserve capacity bid are somehow controversial in academia. Namely, many existing scenario-based works directly manage generation outputs among all possible scenarios to satisfy scenario-wise operational constraints and minimize the expected production cost, with \cite{2009RuizTPS} and \cite{2011OrenTPS} being classical works and \cite{Cong2022PSCC} being a latest one. Such scenario-wise generation scheduling \textit{implicitly} ensures the system to be operationally flexible. And in that sense, the reserve ancillary service, which is an \textit{explicit} measurement of flexibility from generators, is not directly modeled in these papers, along with the neglect of the associated capacity bid.
    
However, this may not be in line with real-world market practices. Currently, most ISOs organize bid-based reserve markets, like NYISO\cite{NYISOReserveBid} and PJM\cite{PJM}. In fact, to provide reserve, generators need to constantly operate at varying load levels instead of the base-load level that they are designed for, while such constant cycling would result in the heat-rate degradation of certain generators\cite{ReserveCost2013}, e.g., fossil steam units and combined cycle units according to PJM's Manual 15 on cost development \cite{PJMm15}. In that manual, the cost-based reserve bid calculations for these generators are exactly built on their possible degrading losses.

Therefore, in this paper, we adopt a practical model that incorporates generators' reserve capacity bids, and make some efforts on reserve pricing to ensure generators' bid-in cost recovery. We hope our work can bridge the gap between those existing energy-only stochastic optimization efforts from academia and current industrial practices on reserve markets.

In assumption (ii), we regard renewable generations as negative loads. In fact, they are also potential reserve providers with stochastic generation upper bounds, and there are several existing pilot explorations on integrating them into AS markets in the U.S., see \cite{2021ReportRenewInMarket} for a survey. However, their market participation models are still unclear as discussed in \cite{2021ReportRenewInMarket} and cannot be fully addressed as a part of this paper. Therefore, we will leave it for our future work.

Assumption (iii) is standard and consistent with current electricity market design.

In assumption (iv), we ignore possible generator outages. Actually, they can be incorporated into the proposed dispatch and pricing through a similar procedure in our previous work\cite{shi2020scenariooriented}. However, such extensions may bring in the \textit{heterogeneity} of energy and reserve from generators for their \textit{differentiated} outage probabilities. Since our main focus for this paper is on their ramping-induced heterogeneity, we therefore do not consider generator outages to make our work more concentrated.

In assumption (iv), we also use reserve to deal with both possible outages and load/renewable forecast errors. In practice, these two kinds of uncertainties are generally handled by different reserve products, i.e., regulating reserve is often utilized to handle load/renewable forecast errors, while spinning reserve is the main tool to cope with contingencies. Currently, these two different reserve products are both provided mainly by generators. However, considering the enormous reserve need brought by increasing penetration of renewable generations and the fact that other potential suppliers (like virtual power plants) are not yet participating in the reserve market on a large scale, counting on generators to separately set aside generation capacities and ramping capabilities respectively for these two reserve products would place heavy burdens on their operations. Therefore, we attempt to merge these two reserve products, and explore the market design to use one single reserve product to handle both possible outages and load/renewable forecast errors.

Meanwhile, we do realize that the control schemes of regulating reserve and spinning reserve are different in industrial practice, which is another key problem in merging these two reserve products apart from the market design issue that we focus on in this paper. Due to the page limit, this problem cannot be addressed as a part of this paper, which is a possible limitation of this work and will be left for our future research.

In assumption (v), we regard the probabilities of possible scenarios as given parameters. How to obtain them is indeed out of the scope of this paper, but here we roughly mention some existing related research for reference. Namely, to obtain the probabilities of mixed scenarios covering both load/renewable uncertainties and contingencies, we can assume that contingencies and load/renewable uncertainties are independent as in \cite{Conejo2013TPS,Hug2020TPS}. Then we can separately calculate the probabilities of load/renewable scenarios and contingency scenarios, and combine them to get the probabilities of mixed ones. How to compute the probabilities of contingency scenarios has been discussed in \cite{2010ConejoBook} in detail. Meanwhile, based on the probability distributions of renewable generations, authors of \cite{2011OrenTPS} have proposed a novel methodology to select typical renewable scenarios and weight them with proper probabilities, which can also be applied for load uncertainties.

\subsection{Key points related to market implementations}

Besides the review on underlying assumptions, some key technical points on implementing the proposed framework into current market practices are also discussed in detail. 

For regional markets with two-settlement structures, the proposed framework can be regarded as an intra-day market between the day-ahead scheduling and the real-time 5-minute economic dispatch. Specifically, if i) the time duration of each interval is 15 minutes, ii) the energy clearing result $g^{*}_{it},\pi^g_{it}$ for the immediate interval is only financially binding, and iii) the reserve result $r^{U*}_{it},r^{D*}_{it},\pi^U_{it},\pi^D_{it}$ is both financially and physically binding, then the proposed market clearing is almost consistent with CAISO's FMM module\cite{CAISOtariff}, with differences in detailed timeline settings and optimization techniques. 

At the same time, we also realize that some regions are still exploring the deregulation of electric utilities, then our work provides an ad hoc solution to their real-time market constructions, and the results of energy and reserve for the immediate interval should be both financially and physically binding in this case.

A closely related question is how to settle possible generation re-dispatch. If the proposed framework is inserted into any two-settlement system as an intermediate step, then the pricing of re-dispatch should follow the rules of the subsequent real-time market. For example, in ERCOT, re-dispatch is settled at the real-time LMP\cite{ercot}. Meanwhile, for cases where the proposed framework acts as the terminal market-clearing process before real-time operations, the pay-as-bid mechanism can be adopted instead of organizing another round of marginal pricing-based auctions to reduce computational burdens.

When discussing possible methods of re-dispatch settlement, it is pertinent to note that the dominant proportion of generation reserve and re-dispatch originates from load/renewable uncertainties, thus how to charge these uncertainty sources is a non-trivial question for the design of stochastic electricity markets. In this paper, we let loads pay for their possible deviations from their forecast powers in all scenarios as in (\ref{Load fluctuation payment}). In contrast, another method is to let loads pay only for their actual deviations in a probability-adjusted manner \cite{Kazempour2018}. Note that the scenario-dependent latter may expose both the ISO and customers to significant financial risks. See our previous work \cite{shi2020scenariooriented} for detailed comparisons.

These discussions on detailed market rules are all related to the ISO's surplus from organizing the market. Unfortunately, revenue adequacy of the ISO cannot be guaranteed under the proposed market design, which is a possible limitation of this work. One main reason is that the property typically lies beneath the dispatch and pricing given by one single procurement model, whereas the market-clearing decisions over the scheduling period are somehow separately generated by a sequence of look-ahead optimizations in this paper. In general, this is an intractable problem for all rolling-window-based market approaches. Another reason is that the settlement of possible re-dispatch, which naturally influences the ISO's surplus in the look-ahead stage, somehow depends on the result of the subsequent real-time market if the proposed framework performs as an intermediate market step instead of the terminal one. In the real world, the ISO is not likely to be able to obtain perfect oracles on the real-time market outcome from a look-ahead perspective, let alone the fact that the ISO's look-ahead decision itself would also influence the real-time result. Such paradoxes cannot be fully addressed as a part of this paper and will be left for our future work.

\section{Case Study}
In this section, two toy examples were used to show the non-trivial coupling between energy, reserve, and ramp. Meanwhile, simulations on the ISO-NE 8-zone system \cite{ISONEsource1,ISONEsource2} with 76 generators were also performed. Note that all quantities were in MW and prices were in \$/MWh, which were dropped hereafter for simplicity.

\subsection{One-bus toy example}
\begin{figure}[t]
		\centering
	\vspace{-1.3cm}
		\includegraphics[width=3.45in]{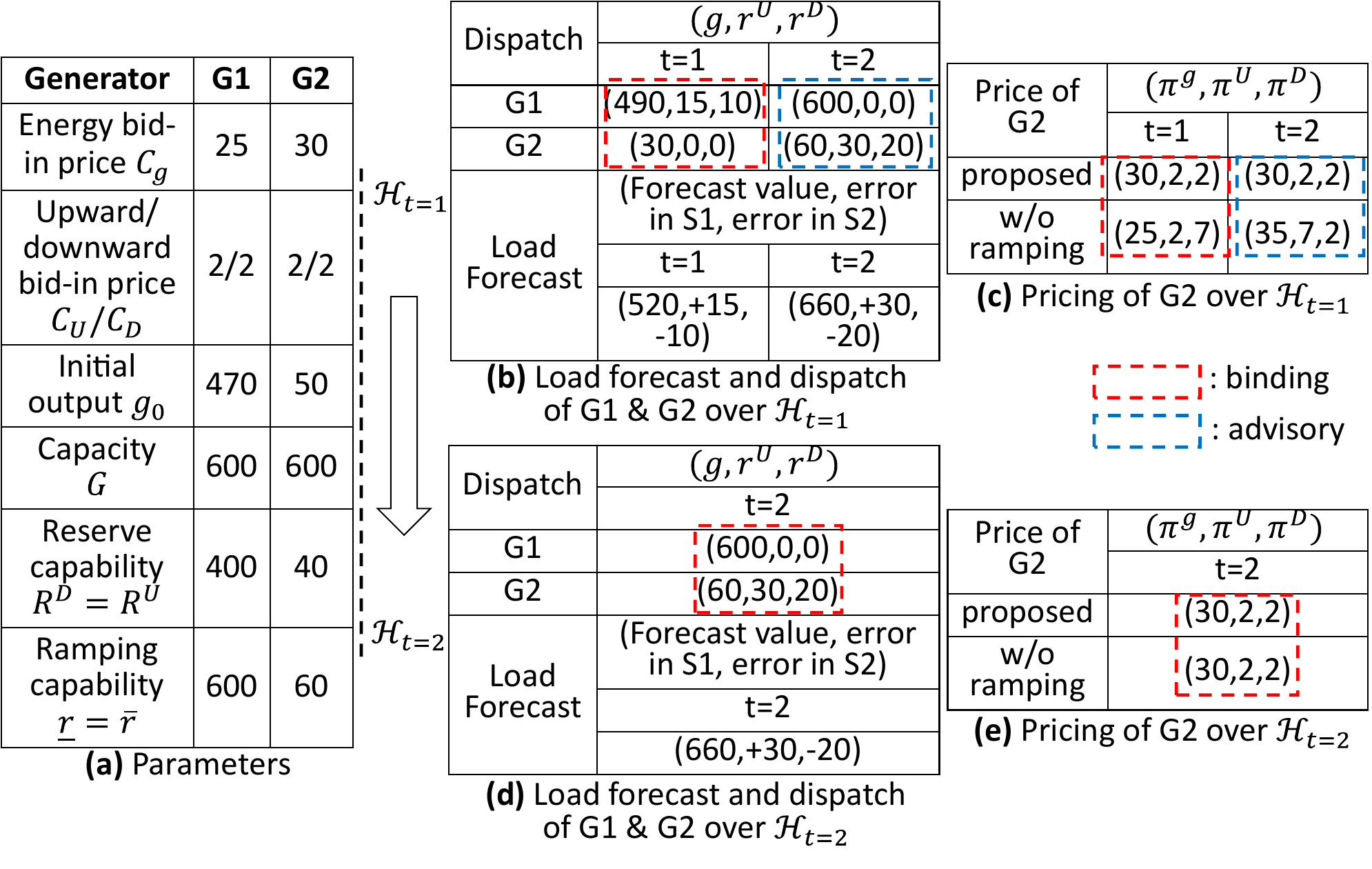}
    \vspace{-0.70cm}
    \captionsetup{font=footnotesize,singlelinecheck=false}
		\caption{(a) Generator parameters, (b) look-ahead load forecast and dispatch over $\mathscr{H}_{t=1}$, (c) look-ahead pricing of G2 over $\mathscr{H}_{t=1}$, (d) load forecast and dispatch over $\mathscr{H}_{t=2}$, and (e) pricing of G2 over $\mathscr{H}_{t=2}$ of the one-bus case}
		\label{fig:one bus example}
    \vspace{-0.7cm}
	\end{figure}

We first studied the rolling-window dispatch and pricing for a one-bus case with two generators G1, G2 and one load considering two 15-minute intervals. For G1 and G2, their parameters were given in Fig. \ref{fig:one bus example}(a). To clear the market for interval $1$, a co-optimization was conducted over the look-ahead window $\mathscr{H}_{t=1}=\{1,2\}$ based on the load forecast in Fig. \ref{fig:one bus example}(b), where the load power uncertainty over $\mathscr{H}_{t=1}$ was captured via two scenarios 1 and 2 apart from the system nominal operation with different forecast errors and probabilities 0.1 and 0.15, respectively. Then the binding result for interval 1 was inputted into the co-optimization over the shrinking look-ahead window $\mathscr{H}_{t=2}=\{2\}$ to clear the market for interval $2$ based on unchanged load forecast in Fig. \ref{fig:one bus example}(d).

From the look-ahead dispatch result over $\mathscr{H}_{t=1}$ in Fig. \ref{fig:one bus example}(b), we could observe that to satisfy the 690MW load demand in scenario 1 at interval $2$, G1 would reach its upper bound 600MW and leave the rest 90MW to G2, including 60MW energy and 30MW upward reserve of G2. The deployment of 30MW upward reserve at interval $2$ would occupy a half of G2's ramping capability from interval $1$ to interval $2$, and left only 30MW ramping capability for its scheduled generation ramp between these two intervals, which forced G2 to have an output of 30MW at interval $1$ while the cheaper G1 was still available to generate more. 

As discussed above, because of its reserve provision, G2's ramping was limited although its scheduled energy difference between interval $1$ and interval $2$ was smaller than its ramping rate, and this might further result in its missing-money problem under the traditional pricing but would be overcome under the proposed pricing. Namely, the proposed pricing was compared with the traditional single-interval scenario-wise LMP and reserve pricing in \cite{shi2020scenariooriented}, which only considered the marginal cost at any single interval but did not incorporate the temporal effect into the pricing, and would be referred to as the \textbf{w/o ramping} benchmark in Fig. \ref{fig:one bus example} and other followed figures for brevity. Under the benchmark pricing, with the prices in the last row of Fig. \ref{fig:one bus example}(c) for interval $1$, G2 suffered \$37.5 loss at 15-minute interval $1$ considering its bid-in costs in Fig. \ref{fig:one bus example}(a), and had zero profit at interval $2$ with the prices in the last row of Fig. \ref{fig:one bus example}(e) for interval $2$, leading to its inadequate payment of \$37.5 over the 2-interval scheduling period. In contrast, as in the second row of Fig. \ref{fig:one bus example}(c) and Fig. \ref{fig:one bus example}(e), prices of G2 assigned by the proposed pricing were no less than its bid-in costs in Fig. \ref{fig:one bus example}(a) at both intervals, avoiding G2's underpayment.

\subsection{Two-bus toy example}
\begin{figure}[t]
		\centering
	\vspace{-1.32cm}
		\includegraphics[width=3.45in]{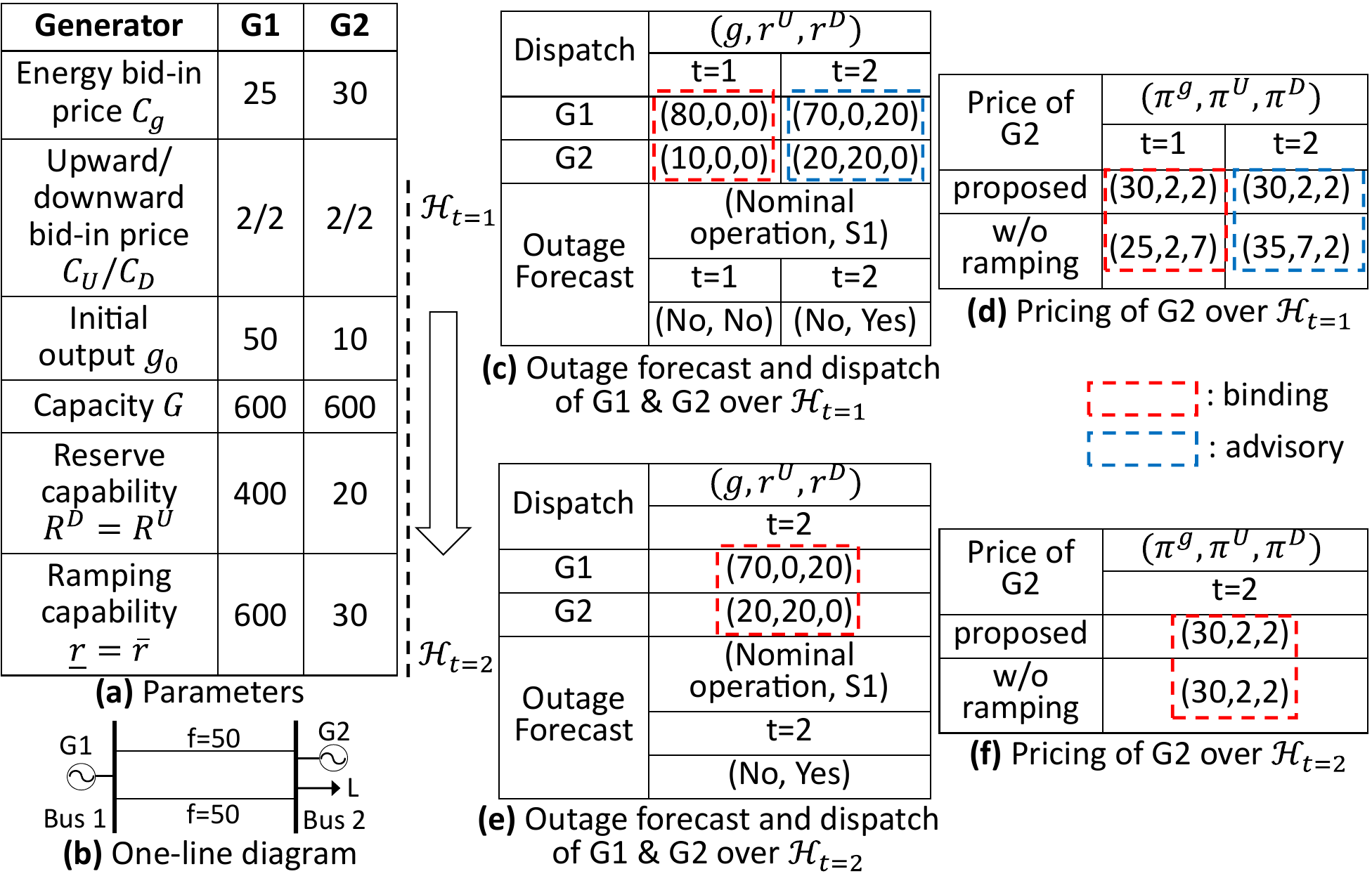}
    \vspace{-0.75cm}
    \captionsetup{font=footnotesize,singlelinecheck=false}
		\caption{(a) Generator parameters, (b) one-line diagram, (c) look-ahead outage forecast and dispatch over $\mathscr{H}_{t=1}$, (d) look-ahead pricing of G2 over $\mathscr{H}_{t=1}$, (e) outage forecast and dispatch over $\mathscr{H}_{t=2}$, and (f) pricing of G2 over $\mathscr{H}_{t=2}$ of the two-bus case}
		\label{fig:two bus example}
    \vspace{-0.8cm}
	\end{figure}

In the previous one-bus example, generators' reserve for load uncertainties resulted in limited ramping and further led to inadequate payment. While in this example, we analyzed the rolling-window dispatch of a two-bus case with simple networks for two 15-minute intervals to show that reserve prepared for outages could also bring about similar problems. As shown from the one-line diagram in Fig. \ref{fig:two bus example}(b), we considered one \textit{constant} 90MW load at bus 2 and two generators G1 and G2 at bus 1 and bus 2, respectively, with their parameters in Fig. \ref{fig:two bus example}(a). These two buses were linked by a transmission line of two identical branches, each of which had a rating of 50MW. Over the first look-ahead window $\mathscr{H}_{t=1}=\{1,2\}$, we considered one contingency scenario with a probability of 0.1, which characterized the outage for one of the two branches at interval $2$ apart from the contingency-absent nominal operation as shown in Fig. \ref{fig:two bus example}(c), and the second look-ahead window $\mathscr{H}_{t=2}=\{2\}$ inherited such scenario setting as in Fig. \ref{fig:two bus example}(e).

Reserve-induced binding ramping also appeared in this example. Namely, from the look-ahead dispatch result over $\mathscr{H}_{t=1}$ in Fig. \ref{fig:two bus example}(c), a counter-intuitive phenomenon was that while the cheaper G1 was still able to generate more and the transmission line still had available capacities to deliver G1's generation to bus 2 at interval $1$, the more expensive G2, however, had a non-zero output of 10MW at interval $1$. The reason was that with the outage of one branch at interval $2$ in the contingency scenario, G1 could only generate 50MW in that scenario limited by the halved post-contingency line capacities, leaving the remaining 40MW load demand to G2. This required G2 to provide 20MW energy and 20MW upward reserve at interval $2$, and 20MW reserve deployment would occupy 2/3 of G2's ramping capability from interval $1$ to interval $2$, forcing G2 to generate 10MW at interval $1$.

As noted above, G2's reserve caused by congestion in the outage scenario induced limited ramping even if its scheduled energy difference between these two intervals was only 1/3 of its ramping rate, and this might further lead to its inadequate payment. Namely, considering G2's bid-in costs in Fig. \ref{fig:two bus example}(a), the prices given by the single-interval scenario-wise pricing in the last row of Fig. \ref{fig:two bus example}(d) for interval $1$ resulted in \$12.5 deficit of G2 at 15-minute interval $1$, as well as zero profit at interval $2$ with the prices in the last row of Fig. \ref{fig:two bus example}(f) for interval $2$, leading to G2's \$12.5 underpayment over the 2-interval scheduling period. In contrast, by comparing the second row of Fig. \ref{fig:two bus example}(d) and Fig. \ref{fig:two bus example}(f) with G2's bid-in costs in Fig. \ref{fig:two bus example}(a), we could observe that G2 was not underpaid under the proposed pricing.

\subsection{ISO-NE 8-zone case}

\subsubsection{Test settings}
\begin{figure}[t]
		\centering
	\vspace{-1.4cm}
		\includegraphics[width=3.4in]{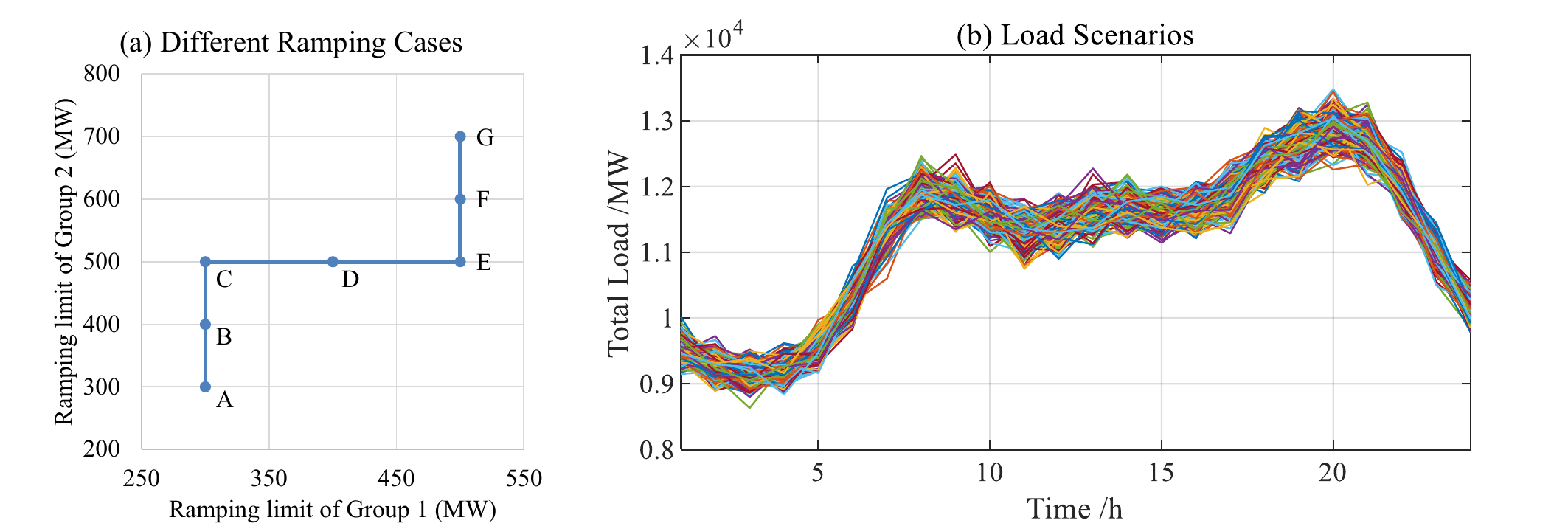}
    \vspace{-0.15cm}
\captionsetup{font=footnotesize,singlelinecheck=false}
		\caption{(a) Ramping cases and (b) load paths of ISO-NE 8-zone case}
   \label{fig:load scenarios and ramping cases}
    \vspace{-0.80cm}
    \end{figure}

Simulations were also performed on the 8-zone ISO-NE system for 300 independent scheduling days under different ramping cases. The topology profile and generator parameters were originally from \cite{ISONEsource1,ISONEsource2}. In this paper, the capacities of all transmission lines were set as 900MW, with a short-term exceeding rate of 1.1 after uncertainty realizations and re-adjustments in all scenarios. At the same time, the bid-in costs of upward and downward reserve were set as 1/5 of energy bid-in costs. Under these settings, the proposed market mechanism was tested under varying ramping capability levels of generators. In the ISO-NE 8-zone case, there were originally two groups of generators with distinguished ramping limits, i.e., Group 1 including G1-G31, and Group 2 including G32-G76. For these two groups of generators, we defined seven \textbf{ramping cases A-G}: for the most relaxed ramping case G, all Group-1 generators' ramping limits were set at 500MW, and all Group-2 generators' ramping limits were 700MW; for the most stringent A, both Group-1 and Group-2 generators' ramping limits were set at 300MW; A, G, and five other ramping cases B-F were depicted in Fig. \ref{fig:load scenarios and ramping cases}(a).

After giving seven ramping cases, we then defined 300 \textbf{independent} scheduling days. On one hand, with the hourly zonal load profile in ISO-NE as the average load path\cite{ISONEload}, 300 load realizations for these scheduling days were generated with a standard deviation of 4\% and presented in Fig. \ref{fig:load scenarios and ramping cases}(b). On the other hand, for the equipment status in these scheduling days, we considered possible outage of Line 11 between zone CT and zone RI and assumed its outage to follow a Markov process with a well-functioned initial state at the beginning of each scheduling day, an outage rate of 1\% per hour, and a zero repair rate. Based on this model, 300 realizations of Line 11's 24-hour status were generated for these scheduling days.


After defining 300 independent scheduling days, for each of them, simulations were done via the proposed rolling-window dispatch, where each look-ahead window covered 4 hourly intervals. For the forecast model, when the ISO stood at $t-1$ and conducted the look-ahead forecast over $\mathscr{H}_t$, the forecast error of any load at interval $t+k-1$ was assumed to follow the normal distribution with a zero mean and a 0.036$k$\% variance, and the ISO's knowledge on Line 11 was limited to its status at $t-1$ and its Markov model described above. To consider these uncertainties, 50 scenarios of forecast errors and outages were generated to drive each look-ahead dispatch.

\subsubsection{System cost and ISO's surplus}
\begin{figure}[t]
\centering
\vspace{-1.2cm}
\label{fig:costs and surplus}
\includegraphics[width=3.4in]{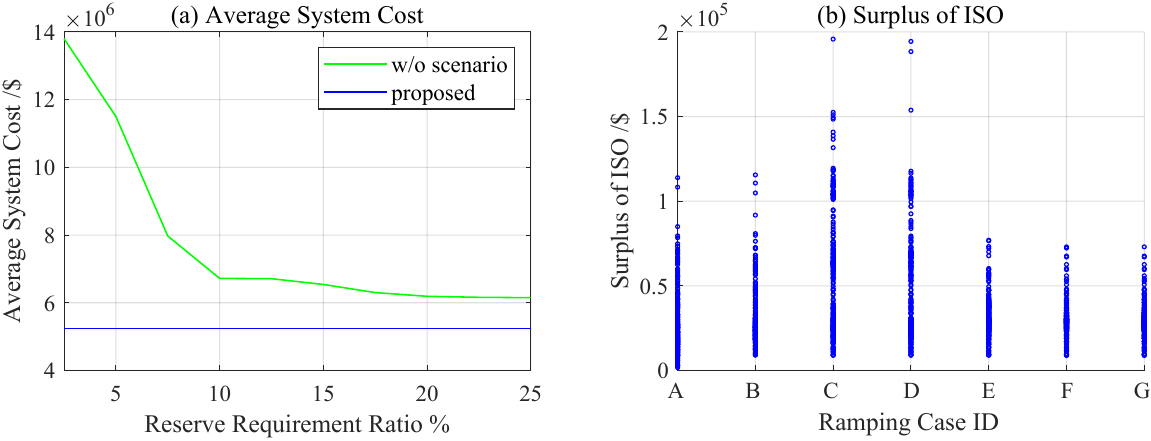}
\vspace{-0.20cm}
\captionsetup{font=footnotesize,singlelinecheck=false}
\caption{(a) Average system costs under ramping case A and (b) the ISO's surplus in all scheduling days under ramping cases A-G}
\label{fig:cost and surplus}
\vspace{-0.10cm}
\end{figure}

We first compared the average system cost of the proposed model among these scheduling days under ramping case A with that of the deterministic model formulated in the Appendix E, which was referred to as the \textbf{w/o scenario} benchmark in Fig. \ref{fig:cost and surplus}(a) and other followed figures. Namely, for any interval in any scheduling day, we first calculated the binding energy and reserve results and clearing costs with the proposed model, and then calculated the re-adjustment costs of re-dispatch and load shedding to handle actual load forecast errors and outages at that interval. By adding them up, we could obtain the system total cost of the proposed model at that interval. Then we repeated that process for the deterministic model under different reserve requirement parameters, i.e., different ratios of system total loads. The average result among all scheduling days was presented in Fig. \ref{fig:cost and surplus}(a), where the green curve represented the average system total cost of the deterministic model under different reserve requirements, and the blue line represented the result of the proposed model. We could observe that the proposed model achieved a lower system total cost than the traditional deterministic model in this test.

In addition to the system cost, we also evaluated the ISO's merchandise surplus under the proposed method as in Fig. \ref{fig:cost and surplus}(b), where each blue circle represented the ISO's surplus in one scheduling day under one ramping case. The result in Fig. \ref{fig:cost and surplus}(b) showed that in all scheduling days, the merchandise surplus of the ISO stayed non-negative under all ramping cases, validating the ISO's revenue adequacy in this example.

\subsubsection{Incentive compatibility}
\begin{figure}[t]
\centering
\vspace{-0.3cm}
\includegraphics[width=3.4in]{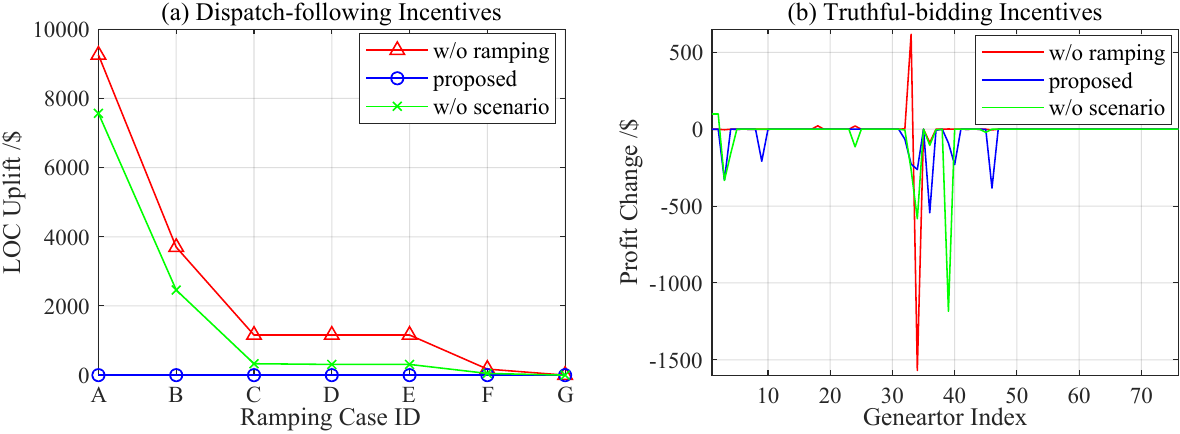}
\vspace{-0.20cm}
\captionsetup{font=footnotesize,singlelinecheck=false}
\caption{(a) Average LOC uplifts under ramping cases A-G and (b) average profit changes of generators from truthful ramping revealing to withholding 20\% ramping under ramping case A}
\label{fig:incentives}
\vspace{-0.70cm}
\end{figure}

Next we verified dispatch-following incentives of generators, which were evaluated via the LOC uplift calculated by (\ref{LOC calculation}). Specifically, under ramping cases A-G, the average uplift of the proposed pricing among these scheduling days was presented in Fig. \ref{fig:incentives}(a) as the blue curve, and was compared with the results of two benchmarks: (i) the red curve represented the result of the single-interval scenario-wise LMP and reserve pricing obtained from the proposed scenario-based model, and (ii) the green curve represented the result of the deterministic LMP and reserve pricing obtained from the deterministic model in the Appendix E. We could observe that the uplift under the proposed pricing was constantly at zero under different ramping cases, which validated dispatch-following incentives of the proposed pricing. On the contrary, the uplift was necessarily non-zero under both benchmarks, and would grow higher with the ramping becoming more stringent.

Meanwhile, to examine generators' incentives to reveal their truthful ramping, under ramping case A, we calculated the average profit change of each generator from truthful ramping revealing to withholding its ramping capability by 20\% among these scheduling days and presented the result in Fig. \ref{fig:incentives}(b), where the blue curve represented the result under the proposed pricing, while the red and green curves respectively corresponded to benchmarks (i) and (ii) described above. We could see that under both benchmarks, some generators could benefit from revealing false ramping, e.g., G1/G2 under the deterministic LMP and reserve pricing, and G24/G33 under the single-interval scenario-wise pricing. On the contrary, under the proposed pricing, no generator could benefit from hiding true ramping, validating their ramping-revealing incentives.

\section{Conclusion}
    This paper provides a solution to the intra-day rolling-window joint dispatch and pricing of energy and reserve, which is necessary for the accommodation of increasing volatile and uncertain renewables but has not been properly addressed yet in both industry and academia. A novel rolling-window energy-reserve co-optimization has been proposed to coordinate inter-temporal ramp with intra-temporal reserve and minimize the expected system total cost considering scenario trajectories of possible outages and load/renewable deviations.
	
    Based on the proposed model, marginal prices of energy and reserve have been derived, which incorporate the Lagrangian multipliers of generators' private ramping limits to eliminate their possible ramping-induced opportunity costs or arbitrages. Indicated by the proposed pricing, generation resources at the same bus are heterogeneous for their differentiated ramping capabilities, and reserve resources are further heterogeneous for their differentiated re-dispatch marginal costs. Based on the proposed dispatch and pricing, dispatch-following and truthful-bidding incentives of generators have been proved, and revenue adequacy of the ISO has been numerically validated in the case study. In future studies, more efforts will be made on the reserve market design with diversified participants, like renewables and energy storage systems.

    \bibliographystyle{ieeetr}    
	\bibliography{lib_test}	
	

	\appendices
    
    \section{Notations and Nomenclatures}
    \label{sec:Notations}
    The notations and nomenclatures are presented in Table \ref{symbols}.

    {\small
\begin{table}[H]
\vspace{-0.5cm}
\begin{center}
\caption{\small List of Major Symbols Used}
\label{symbols}
\begin{tabular}{|ll|}
\hline
$\bm{f}$: & line capacities\\
$C^g_{it},C^U_{it},C^D_{it}$: & generator $i$'s bid-in prices of energy, \\&upward and downward reserve at $t$\\
$C^L_{lt}$: & load $l$'s shedding price at $t$\\
$\hat{d}_{lt}$: & load $l$'s forecast power at $t$\\
$\delta g^U_{it,s},\delta g^D_{it,s}$: & upward and downward re-dispatch of \\ &   generator $i$ at $t$ in scenario $s$\\
$\delta d_{lt,s}$: & load $l$'s shedding  at $t$ in scenario $s$\\
$\epsilon_s$: & occurrence probability of scenario $s$\\
$F^t$: & expected system total cost over $\mathscr{H}_t$\\
$F^t_{-it}$: & $F^t$ excluding the energy and reserve \\& bid-in cost of generator $i$ at $t$\\
$g_{it},r^U_{it},r^D_{it}$: & energy, upward and downward reserve\\& provided by generator $i$ at $t$\\
$\mathscr{H}$: & scheduling period $\mathscr{H} = \{1,\cdots,T\}$\\
$\mathscr{H}_t$: & look-ahead window starting from $t$\\
$\widetilde{\bm{P}}_i$: & rolling-window dispatch over $\mathscr{H}$\\
$\pi^g_{it},\pi^d_{it}$: & energy marginal prices of generator $i$\\& and load $l$ at $t$\\
$\pi^U_{it},\pi^D_{it}$: & upward and downward reserve marginal\\& prices of generator $i$ at $t$\\
$\widetilde{\bm{\pi}}_i$: & rolling-window pricing over $\mathscr{H}$\\
$\overline{r}_{it},\underline{r}_{it}$: & upward and downward \\& ramping capabilities of generator $i$ at $t$\\
$\bm{S}_{G},\bm{S}_{D}$: &  shift-factor matrices of generators/loads\\
$\hat{\xi}_{lt,s}$: & load $l$'s forecast error at $t$ in scenario $s$\\\hline
\end{tabular}
\end{center}
\vspace{-0.80cm}
\end{table}
}

	\section{Proof of Dispatch-following Incentives}
	\label{Sec:Proof of dispatch-following incentives}
	We first consider some KKT conditions of the proposed look-ahead co-optimization $F^t(\cdot)$ over $\mathscr{H}_t$ as follows:
	
	\begin{align}
	\label{Clearing KKT energy 0}
	\frac{\partial L^*_t}{\partial g_{it}}=&\!-\!(\lambda^*_{t,0}\!-\!\bm{S}^T_G(\cdot,m_i)\bm{\phi}^*_{t,0})\!-\!\!\sum\limits_{s \in \mathcal{S}_{t}}(\lambda^*_{t,s}\!-\!\bm{S}^T_{G,t,s}(\cdot,m_i)\bm{\phi}^*_{t,s})\notag\\
	&-\!\Big(\overline{\mu}^*_{it}\!-\!\underline{\mu}^*_{it}\Big)\!+\!\Big(\overline{\mu}^*_{i(t-1)}\!-\!\underline{\mu}^*_{i(t-1)}\Big)\!+\!C^g_{it}\!+\!\overline{\upsilon}^*_{it}\!-\!\underline{\upsilon}^*_{it}\notag\\
	=&0,\\
	\label{Clearing KKT up reserve 0}
	\frac{\partial L^*_t}{\partial r^{U}_{it}}=&-\sum\limits_{s \in \mathcal{S}_{t}} \overline{\alpha}^*_{it,s}+\overline{\mu}^*_{i(t-1)}+\underline{\mu}^*_{it}+C^U_{it}+\overline{\upsilon}^*_{it}-\underline{\rho}^{U*}_{it}+\overline{\rho}^{U*}_{it}\notag\\
	=&0,\\
	\label{Clearing KKT down reserve 0}
	\frac{\partial L^*_t}{\partial r^{D}_{it}}=&-\sum\limits_{s \in \mathcal{S}_{t}} \overline{\beta}^*_{it,s}+\underline{\mu}^*_{i(t-1)}+\overline{\mu}^*_{it}+C^D_{it}+\underline{\upsilon}^*_{it}-\underline{\rho}^{D*}_{it}+\overline{\rho}^{D*}_{it}\notag\\
	=&0,
	\end{align}
	where $L^*_t$ represents the optimal Lagrangian of $F^t(\cdot)$. With the pricing formulations in (\ref{Gen energy price})-(\ref{Gen down res price}) of $\pi^g_{it},\pi^U_{it},\pi^D_{it}$, these KKT conditions can be further transformed into
	\begin{flalign}
	\label{Clearing KKT energy}
	&\frac{\partial L^*_t}{\partial g_{it}}=-\pi^g_{it}+C^g_{it}+\overline{\upsilon}^*_{it}-\underline{\upsilon}^*_{it}=0,\\&
	\label{Clearing KKT up reserve}
	\frac{\partial L^*_t}{\partial r^{U}_{it}}=-\pi^U_{it}+C^U_{it}+\overline{\upsilon}^*_{it}-\underline{\rho}^{U*}_{it}+\overline{\rho}^{U*}_{it}=0,\\&
	\label{Clearing KKT down reserve}
	\frac{\partial L^*_t}{\partial r^{D}_{it}}=-\pi^D_{it}+C^D_{it}+\underline{\upsilon}^*_{it}-\underline{\rho}^{D*}_{it}+\overline{\rho}^{D*}_{it}=0.&
	\end{flalign}
	We next consider the case where each generator $i$ self-schedules its energy and reserve provision at $t$ to maximize its profit at $t$ with given prices $\bm{\pi}_{it}=\{\pi^g_{it},\pi^U_{it},\pi^D_{it}\}$ from the ISO for $t$, and write the corresponding model as:
	\begin{flalign}
	\label{obj_Exante_IR}
	&Q_{it}(\widetilde{\bm{\pi}}_{it})\!=\! \big(\!\pi^g_{it}g_{it}\!+\!\pi^U_{it}\!r^U_{it}\!+\!\pi^D_{it}\!r^D_{it}\big)\!-\! \big(C^g_{it}g_{it}\!+\!C^U_{it}\!r^U_{it}\!+\!C^D_{it}\!r^D_{it}\!\big),\\
    &\underset{\{g_{it},r^U_{it},r^D_{it}\}}{\rm maximize}\:\: Q_{it}(\widetilde{\bm{\pi}}_{it}),\: \mbox{subject to:}&\notag\\
	\label{binding capacity 2}	
	&(\underline{\upsilon}'_{it},\overline{\upsilon}'_{it}):\underline{G}_{it} \leq  g_{it} - r^D_{it},g_{it} + r^U_{it}  \leq \overline{G}_{it},&\\
	\label{binding reserve capability 2}
    &(\underline{\rho}^{D'}_{it},\overline{\rho}^{D'}_{it},\underline{\rho}^{U'}_{it},\overline{\rho}^{U'}_{it}):0 \leq r^D_{it} \leq R^D_{it},0 \leq r^U_{it} \leq R^U_{it},&\\
    \label{binding inter-temporal up 2}
	&(\underline{\mu}'_{i(t-1)}): -\underline{r}_{it} \leq g_{it}-g_{i(t-1)}-r^D_{it}-r^U_{i(t-1)},&\\
	\label{binding inter-temporal down 2}
	&(\overline{\mu}'_{i(t-1)}):g_{it}-g_{i(t-1)}+r^U_{it}+r^D_{i(t-1)} \leq \overline{r}_{it}.&
	\end{flalign}
	For the above model, if we set
	\begin{flalign}
    &\underline{\mu}'_{i(t-1)}=0,\overline{\mu}'_{i(t-1)}=0,\underline{\rho}^{D'}_{it}=\underline{\rho}^{D*}_{it},\overline{\rho}^{D'}_{it}=\overline{\rho}^{D*}_{it},\notag&\\
    \label{Lagrangian equal 1}
    &\underline{\rho}^{U'}_{it}=\underline{\rho}^{U*}_{it},\overline{\rho}^{U'}_{it}=\overline{\rho}^{U*}_{it},\underline{\upsilon}'_{it}=\underline{\upsilon}^*_{it},\overline{\upsilon}'_{it}=\overline{\upsilon}^*_{it}&
	\end{flalign}
	where $\underline{\rho}^{D*}_{it},\overline{\rho}^{D*}_{it},\underline{\rho}^{U*}_{it},\overline{\rho}^{U*}_{it},\underline{\upsilon}^*_{it},\overline{\upsilon}^*_{it}$ are optimally solved from the proposed co-optimization $F^t(\cdot)$ over $\mathscr{H}_{t}$, then based on (\ref{Clearing KKT energy})-(\ref{Clearing KKT down reserve}), we can observe that this set of dual variables in (\ref{Lagrangian equal 1}) and the set of primal variables $\{g^*_{it},r^{U*}_{it},r^{D*}_{it}\}$ solved from $F^t(\cdot)$ satisfy the KKT conditions of the model in (\ref{obj_Exante_IR})-(\ref{binding inter-temporal down 2}), which indicates that
	\begin{equation}
    \{g^*_{it},r^{U*}_{it},r^{D*}_{it}\}=\underset{\{g_{it},r^{U}_{it},r^{D}_{it}\}}{\operatorname{argmax}}\{Q_{it}(\widetilde{\bm{\pi}}_{it})|(\ref{binding capacity 2})-(\ref{binding inter-temporal down 2})\}.\notag
    \end{equation}
    
    Furthermore, from (\ref{Lagrangian equal 1}), we can observe that $\{g^*_{it},r^{U*}_{it},r^{D*}_{it}\}$ optimizes $Q_{it}(\widetilde{\bm{\pi}}_{it})$ with the Lagrangian multipliers of ramping constraints (\ref{binding inter-temporal up 2})-(\ref{binding inter-temporal down 2}) being zero, what means that $\{g^*_{it},r^{U*}_{it},r^{D*}_{it}\}$ also optimizes $Q_{it}(\widetilde{\bm{\pi}}_{it})$ for the ramp-unconstrained case, i.e.,
	\begin{equation}
    \{g^*_{it},r^{U*}_{it},r^{D*}_{it}\}=\underset{\{g_{it},r^{U}_{it},r^{D}_{it}\}}{\operatorname{argmax}}\{Q_{it}(\widetilde{\bm{\pi}}_{it})|(\ref{binding capacity 2})-(\ref{binding reserve capability 2})\}.\notag
    \end{equation}    
    This can be easily augmented to the ramp-unconstrained multi-interval profit-maximization over $\mathscr{H}$ as
	\begin{equation}
    \{g^*_{it},r^{U*}_{it},r^{D*}_{it}, \forall t \!\in\! \mathscr{H}\!\}\!=\!\underset{\{\!g_{it},r^{U}_{it},r^{D}_{it}, \forall t \in \!\mathscr{H\!}\}}{\operatorname{argmax}}\!\{Q_i(\widetilde{\bm{\pi}}_i)|(\ref{binding capacity 3})-(\ref{reserve capability 3})\}.\notag
    \end{equation}
    Meanwhile, since the ramping constraints of generators are indeed incorporated into every $F^t(\cdot)$ for $t \in \mathscr{H}$, then it is obvious that $\{g^*_{it},r^{U*}_{it},r^{D*}_{it}, \forall t \in \mathscr{H}\}$ given by the ISO definitely satisfies generators' ramping constraints over $\mathscr{H}$, which indicates that $\{g^*_{it},r^{U*}_{it},r^{D*}_{it}, \forall t \in \mathscr{H}\}$ also optimizes $Q_i(\widetilde{\bm{\pi}}_i)$ for the ramp-constrained case, i.e.,
    \begin{equation}
    \{g^*_{it},r^{U*}_{it},r^{D*}_{it}, \forall t \!\in\! \mathscr{H}\!\}\!=\!\underset{\{\!g_{it},r^{U}_{it},r^{D}_{it}, \forall t \in \!\mathscr{H}\!\}}{\operatorname{argmax}}\!\{Q_i(\widetilde{\bm{\pi}}_i)|(\ref{binding capacity 3})-(\ref{binding inter-temporal down 3})\},\notag
    \end{equation}
    proving Theorem 1.

    \section{Proof of Cost Recovery}
    \label{Sec: Proof of Cost Recovery}
    The proof of cost recovery is quite intuitive. Namely, With the additional zero-lower-bound assumption, consider any generator $i$. A feasible solution to its multi-interval profit-maximization in (\ref{obj_Expost_IR_multi})-(\ref{binding inter-temporal down 3}) is to provide zero energy and reserve over $\mathscr{H}$ and earn a zero bid-in profit over $\mathscr{H}$. At the same time, according to Theorem 1, the optimal solution is to follow the ISO's dispatch of energy and reserve $\widetilde{\bm{P}}_i$ over $\mathscr{H}$. Therefore, by following $\widetilde{\bm{P}}_i$ over $\mathscr{H}$, generator $i$ can earn an overall bid-in profit that is no less than zero, proving its cost recovery.

    \section{Proof of Truthful-bidding Incentives}
    \label{Sec: Proof of Truthful-bidding Incentives}
    We first revisit the strategic bidding model of any generator $i$ given in (\ref{obj_Exante_bidding})-(\ref{binding inter-temporal down bidding}). The truthful operational parameters of generator $i$ in constraints (\ref{binding capacity bidding})-(\ref{binding inter-temporal down bidding}) actually construct its \textit{truthful feasible dispatch region} over $\mathscr{H}$, which limits generator $i$'s bidding behavior as it wants the rolling-window dispatch of energy and reserve from the ISO to fall into that region, otherwise it would not be physically able to follow. 
    
    Moreover, note that when bidding truthfully, the profit of generator $i$ over $\mathscr{H}$ can be expressed as
    \begin{align}
    \label{Truth_Bid_1}
    \Pi_{i}(\bm{\theta}^*_{i})
    {\color{red}=}\!&\sum_{t \in \mathscr{H}}\!\big(\pi_{it}g^*_{it}(\bm{\theta}^*_{i})\!+\!\pi^U_{it}\!r^{U*}_{it}(\bm{\theta}^*_{i})\!+\!\pi^D_{it}\!r^{D*}_{it}(\bm{\theta}^*_{i})\big)\\
    &-\!\sum_{t \in \mathscr{H}}\big(C^{g*}_{it}g^*_{it}(\bm{\theta}^*_{i})\!+\!C^{U*}_{it}\!r^{U*}_{it}(\bm{\theta}^*_{i})\!+\!C^{D*}_{it}\!r^{D*}_{it}(\bm{\theta}^*_{i})\big),\notag
    \end{align}
    where $\{g^*_{it}(\bm{\theta}^*_{i}),r^{U*}_{it}(\bm{\theta}^*_{i}),r^{D*}_{it}(\bm{\theta}^*_{i}),\forall t \in \mathscr{H}\}$ are the ISO's dispatch quantities over $\mathscr{H}$ with truthful bid $\bm{\theta}^*_{i}$ from generator $i$. From dispatch-following incentives in Theorem 1, we can further interpret that
    \begin{align}
    \Pi_{i}(\bm{\theta}^*_{i})
    =&\sum_{t \in \mathscr{H}}\!\big(\pi_{it}g^*_{it}(\bm{\theta}^*_{i})\!+\!\pi^U_{it}\!r^{U*}_{it}(\bm{\theta}^*_{i})\!+\!\pi^D_{it}\!r^{D*}_{it}(\bm{\theta}^*_{i})\big)\notag\\
    &-\sum_{t \in \mathscr{H}}\big(C^{g*}_{it}g^*_{it}(\bm{\theta}^*_{i})\!+\!C^{U*}_{it}\!r^{U*}_{it}(\bm{\theta}^*_{i})\!+\!C^{D*}_{it}\!r^{D*}_{it}(\bm{\theta}^*_{i})\big)\notag\\
    \label{Truth_Bid_2}
    {\color{red}=}&Q^*_i(\cdot|\bm{\theta}^*_{i}),
    \end{align}
    where $Q^*_i(\cdot|\bm{\theta}^*_{i})$ is the optimal value of generator $i$'s multi-interval profit-maximization (\ref{obj_Expost_IR_multi})-(\ref{binding inter-temporal down 3}) with its cost coefficients and operational parameters being truthful $\bm{\theta}^*_{i}$, which further implies that
        \begin{align}
    \Pi_{i}(\bm{\theta}^*_{i})
    =&\sum_{t \in \mathscr{H}}\!\big(\pi_{it}g^*_{it}(\bm{\theta}^*_{i})\!+\!\pi^U_{it}\!r^{U*}_{it}(\bm{\theta}^*_{i})\!+\!\pi^D_{it}\!r^{D*}_{it}(\bm{\theta}^*_{i})\big)\notag\\
    &-\sum_{t \in \mathscr{H}}\big(C^{g*}_{it}g^*_{it}(\bm{\theta}^*_{i})\!+\!C^{U*}_{it}\!r^{U*}_{it}(\bm{\theta}^*_{i})\!+\!C^{D*}_{it}\!r^{D*}_{it}(\bm{\theta}^*_{i})\big)\notag\\
    =&Q^*_i(\cdot|\bm{\theta}^*_{i})\notag\\
    \label{Truth_Bid_3}
    {\color{red}\geq}\!\!&\sum_{t \in \mathscr{H}}\!\!(\!\pi_{it}g_{it}\!+\!\pi^U_{it}\!r^{U}_{it}\!+\!\pi^D_{it}\!r^{D}_{it}\!-\!C^{g*}_{it}\!g_{it}\!-\!C^{U*}_{it}\!r^{U}_{it}\!-\!C^{D*}_{it}\!r^{D}_{it}\!)
    \end{align}
    for any set of $\{g_{it},r^U_{it},r^D_{it},\forall t \in \mathscr{H}\}$ that lies in generator $i$'s truthful feasible dispatch region. Note that no matter how generator $i$ submits its bid $\bm{\theta}_{i}$, the dispatch over $\mathscr{H}$ that it receives must be within its truthful feasible dispatch region. Therefore, any feasible bid $\bm{\theta}_{i}$ definitely corresponds to a set of $\{g_{it},r^U_{it},r^D_{it},\forall t \in \mathscr{H}\}$ in the truthful feasible dispatch region. Combining this with (\ref{Truth_Bid_3}), we can terminally interpret that
	\begin{equation}
	\Pi_i(\bm{\theta}^*_i)\geq \Pi_i(\bm{\theta}_i),
	\end{equation}
	proving Theorem 2.

	\section{Benchmark: Deterministic Co-optimization}
    \label{sec:deterministic model}
    \begin{flalign}
	\label{obj2}
	&U^t=\sum\limits_{\tau \in \mathscr{H}_t}\sum\limits_{i \in N_G} \Big( C^g_{i\tau}g_{i\tau}+C^U_{i\tau}r^U_{i\tau}+C^D_{i\tau}r^D_{i\tau}\Big),&\\
    &\underset{\{g_{i\tau},r^U_{i\tau},r^D_{i\tau},\forall \tau \in \mathscr{H}_t\}}{\rm minimize} U^t(\cdot),\mbox{subject to $\forall \tau \in \mathscr{H}_t$:} \notag&\\
	&\lambda_{\tau}:\sum\limits_{i \in N_G}\! g_{i\tau}\!=\!\sum\limits_{l \in N_D}\! \hat{d}_{l\tau},&\\
	&\bm{\phi}_{\tau}:\bm{S}_{G}\cdot\bm{g}[\tau]\!-\bm{S}_{D}\cdot\hat{\bm{d}}[\tau]\!\leq\! \bm{f},&\\
	&(\chi^U_\tau,\chi^D_\tau):\sum\limits_{i \in N_G}\! r^U_{i\tau}\!=Req^U_{\tau},\sum\limits_{i \in N_G}\! r^D_{i\tau}\!=Req^D_{\tau},&\\
	&\underline{G}_{i\tau} \leq g_{i\tau}-r^D_{i\tau},g_{i\tau}+r^U_{i\tau} \leq \overline{G}_{i\tau},&\\
	&0 \leq r^D_{i\tau} \leq R^D_{i\tau},0 \leq r^U_{i\tau} \leq R^U_{i\tau},&\\
	&-\underline{r}_{i\tau} \leq g_{i\tau}-g_{i(\tau-1)}-r^D_{i\tau}-r^U_{i(\tau-1)},&\\
	& g_{i\tau}-g_{i(\tau-1)}+r^U_{i\tau}+r^D_{i(\tau-1)} \leq \overline{r}_{i\tau},&
	\end{flalign}
    where $Req^U_{\tau}/Req^D_{\tau}$ are the upward/downward reserve requirement parameters at interval $\tau$. Based on this model, the deterministic LMP of any generator $i$ at $t$ is
 	\begin{equation}
	\eta^g_{it}=\lambda^*_{t}-\bm{S}^T_G(\cdot,m_i)\bm{\phi}^*_{t},
	\end{equation}
    and its deterministic upward and downward reserve pricing at $t$ is
        \begin{equation}
	\eta^U_{it}=\chi^U_t,\eta^D_{it}=\chi^D_t,
	\end{equation}
    respectively.

\end{document}